\documentclass[preprint,showpacs,preprintnumbers,amsmath,amssymb,floatfix]{revtex4}
\usepackage{graphicx}
\usepackage{psfig}
\usepackage{dcolumn}
\usepackage{bm}

\begin{document}

\title{
Magnetic phases and reorientation transitions 
in antiferromagnetically coupled multilayers
}

\author{U.K. R\"o\ss ler}
\thanks
{Corresponding author 
}
\email{u.roessler@ifw-dresden.de}
\author{A.N.\ Bogdanov}
\altaffiliation[Permanent address: ]%
{Donetsk Institute for Physics and Technology,
340114 Donetsk, Ukraine
}
\email{bogdanov@kinetic.ac.donetsk.ua}
\affiliation{
Leibniz-Institut f{\"u}r Festk{\"o}rper- 
und Werkstoffforschung Dresden,\\
Postfach 270116,
D--01171 Dresden, Germany
}%

\date{\today}

\begin{abstract}
{
In antiferromagnetically coupled 
superlattices grown on (001) 
faces of cubic substrates,
e.g. based on materials combinations as
Co/Cu,  Fe/Si, Co/Cr, or Fe/Cr,
the magnetic states evolve 
under competing influence of bilinear and biquadratic
exchange interactions, 
surface-enhanced four-fold in-plane anisotropy,
and specific finite-size effects.
Using phenomenological (micromagnetic) theory,
a comprehensive survey of the magnetic states 
and reorientation transitions 
has been carried out for multilayer systems
with even number of ferromagnetic sub-layers
and magnetizations in the plane.
In two-layer systems ($N=2$) 
the phase diagrams in dependence on
components of the applied field in the plane 
include ``swallow-tail'' type regions of (metastable) multistate 
co-existence and a number of continuous and discontinuous
reorientation transitions induced by radial
and transversal components of the applied field.
In multilayers ($N \ge 4$) noncollinear states are spatially
inhomogeneous with magnetization varying  
across the multilayer stack.  
For weak four-fold anisotropy 
the magnetic states under influence of an applied field
evolve by a
complex continuous reorientation 
into the saturated state.
At higher anisotropy they transform into 
various inhomogeneous and asymmetric structures.
The discontinuous transitions between the magnetic states 
in these two-layers and multilayers are characterized 
by broad ranges of multi-phase coexistence of the (metastable) states
and give rise to specific transitional domain structures.  
%
}
\end{abstract}

\pacs{
75.70.-i,
75.50.Ee, 
75.10.-b 
75.30.Kz 
}

\maketitle

%
\section{Introduction}
Multilayers built from ferromagnetic layers 
with various 
spacer layers include 
a wide variety of magnetic film systems 
that have been intensively studied 
during last years.
Due to remarkable phenomena as giant magnetoresistance,
exchange-spring behaviour and/or exchange bias, 
and surface enhanced magnetic anisotropy,
such nanostructures have already found 
a number of applications and are considered 
as promising candidates for nonvolatile 
magnetic recording media.\cite{Moser02}
On the other hand, nanoscale superlattices and
similar structures provide convenient model systems 
to study different aspects of surface magnetism
and magnetic ordering in confining geometry.

In particular, much attention has been given 
to multilayers composed of antiferromagnetically
coupled ferromagnetic nanolayers.
\cite{Howson94,Wang94,Ounadjela92,Steadman02, 
Hellwig03, Itoh03, PRB03,
Zabel94,Graaf97,Jounge02, Zabel96, Picconatto97,
Hilt99, 
Ustinov01,Chesman98,Lauter02}
Such layered synthetic antiferromagnets
can be separated into two classes according
to the symmetries ruling their magnetic properties:
Superlattices with relatively 
strong uniaxial anisotropies
include low-symmetry multilayers with
effective uniaxial magnetic anisotropy in 
the layer planes, 
e.g. epitaxial systems deposited on (110), (211) 
faces of cubic substrates.\cite{Wang94, Steadman02}
Also multilayer systems with perpendicular anisotropy 
belong to this class.\cite{Hellwig03, Ounadjela92, Itoh03}
Magnetization processes in these nanostructures 
are strongly influenced by the uniaxial anisotropy 
which is responsible for specific phenomena
such  as ``surface spin-flop'' \cite{Wang94,PRB03}
or field-induced metamagnetic jumps.\cite{Hellwig03}
On the other hand, 
the uniaxial anisotropy may be absent 
in layered systems with higher symmetry.
These represent another large and intensively
investigated class of synthetic antiferromagnetic 
nanostructures.
Superlattices with planar magnetization grown 
on (001) faces of cubic substrates, 
e.g. multilayers from materials combinations
as  Co/Cu \cite{Zabel94}, Fe/Si \cite{Graaf97,Jounge02}
Co/Cr, \cite{Zabel96,Picconatto97,Hilt99} or
Fe/Cr, \cite{Ustinov01,Chesman98,Lauter02}
belong to this class.
In the case of weak four-fold anisotropy,
their magnetic properties are mostly determined 
by the interlayer exchange interactions
which may include
an important \textit{biquadratic} contribution.
\cite{Graaf97, Ustinov01, Chesman98, Jounge02} 
Evidence of strong biquadratic exchange interaction
has been given in a number of experimental papers
for Fe/Cr two-layers and multilayers \cite{Chesman98, Ustinov01},
for Fe/Si(001) multilayers \cite{Jounge02}).
Strong effective four-fold anisotropies
have been found in systems such as Co/Cr(001) or Fe/Cr(001).
\cite{Zabel94,Picconatto97, Ustinov01,  Hilt99}
%
%
The competition between magnetic anisotropy,
applied fields and exchange energies may cause 
complicated magnetic effects and processes.
In fact, 
a great number of novel magnetic configurations
and remarkable reorientational effects in external 
fields have been found in such superlattices.
\cite{Zabel94,Graaf97,Ustinov01,Lauter02}
In particular, recent experimental results
using modern depth-resolving techniques 
reveal spatially inhomogeneous 
magnetic structures, e.g. in Fe/Cr(001) superlattices,\cite{Lauter02} 
and specific reorientation effects imposed 
by four-fold planar (tetragonal) anisotropy.\cite{Ustinov01}
The understanding and interpretation 
of the complex magnetization processes found 
in such systems \cite{Temst00, Nagy02, Lauter03}
requires a theoretical underpinning.

Theoretical activity in this field is largely based on analytical
and numerical calculations mostly within phenomenological approaches.
\cite{Dieny90,Nortemann92,
Wang94,Demokritov98, Ustinov01,PRB03}
These studies have demonstrated the general 
validity of the phenomenological models 
to describe the magnetization processes
in antiferromagnetic nanostructures.
%
\cite{Wang94,Zabel94,Lauter02,PRB03}
For the system under discussion
the phenomenological theory has been developed to
describe effects of 90~degree couplings
and a concomitant complex evolution of domain structures
in Fe/Cr/Fe layers.\cite{Hubert98}
Four-fold anisotropy effects 
have been theoretically investigated in Ref.~\onlinecite{Dieny90} 
for sandwich structures with $N$=2 ferromagnetic
layers coupled through a spacer.
%
%
Multilayer systems provide also experimental models 
to study effects of the confining surfaces on antiferromagnetic
structures.\cite{Wang94,Mills99}
In this context, specific inhomogeneous magnetic states
described for theoretical models in Ref.~\onlinecite{Nortemann92},
have recently been observed 
in Fe/Cr superlattices.\cite{Lauter02}
The existing theoretical results, however, 
are restricted to special cases 
and are not sufficient for 
an exhaustive description of the magnetic states 
and field-induced reorientional effects 
observed in recent experiment.
\cite{Zabel94,Graaf97,Jounge02, 
Zabel96,Picconatto97,Hilt99,
Ustinov01, Chesman98, Lauter02}

In this paper we provide a theoretical 
analysis of magnetic states and magnetization processes 
in planar antiferromagnetic superlattices 
with and without four-fold anisotropy 
in magnetic fields applied within the multilayer plane.
In such magnetic superlattices
the exchange interlayer coupling is an oscillatory
function of the spacer thickness.\cite{Gruenberg87, Stiles99}
Depending on the spacer thickness an alternating
sequence of ferromagnetic and antiferromagnetic 
interlayer couplings is realized, 
and, by adjusting the spacer thicknesses,
very different strengths of antiferromagnetic coupling
can be realized.
On the other hand, 
the four-fold anisotropy includes interface-induced 
contributions which implies a strong dependency of the 
effective average anisotropy of each ferromagnetic 
layer on the layer thickness.\cite{Goryunov95}
Thus, the effective magnetic interactions 
can vary in extremely broad ranges for such multilayers 
in dependence on the chosen materials combinations and
thicknesses, see e.g. Ref.~\onlinecite{Zabel94},  where
for a Co/Cu(001) two-layer system with a wedged spacer layer
the ratio between four-fold anisotropy energy 
and the exchange coupling is changed 
by orders of magnitude.
In contrast to bulk planar antiferromagnets,
where an essentially fixed hierarchy for the strengths
of the magnetic interactions holds,\cite{FTT90}
in these artificial antiferromagnetic systems 
the ratios between different magnetic
energies, respectively the phenomenological 
parameters in the magnetic free-energy, 
may assume practically arbitrary values.
Moreover, as the interlayer exchange is weak 
compared to direct exchange interactions, 
the fields to induce spin-reorientation phenomena 
are similarly weak and experimentally accessible.

The rich phase diagrams for these systems
precludes an analysis in all details.
The phase space in terms of the phenomenological parameters 
includes a large variety of different magnetic states
with a corresponding multitude of spontaneous 
and field-induced phase-transformations.
In a first step to such an analysis,
all laterally homogeneous states 
in such multilayers must be found.
They are the building blocks 
for a domain theory.\cite{FTT90, UFN88}
For the case of laterally homogeneous states of 
each ferromagnetic sublayer one has a 
system that behaves like 
an antiferromagnetically coupled 
chain of Stoner-Wohlfarth particles.
This simplified one-dimensional model 
for the behaviour across the multilayer stack 
also yields the limiting case for 
the magnetization processes with maximum hystereses.
Again, a direct analysis of all magnetic states 
even for these one-dimensional models 
yields an intricate succession of 
phase diagrams and magnetization curves.\cite{Dieny90}
In this paper, we avoid the cumbersome task
of listing and classifying \textit{all} 
solutions and transitions. 
Instead, we provide a broad 
physically intuitive picture of the physical effects 
due to the different exchange or anisotropic forces, 
and those imposed by the confining geometry of the system. 
To this end, we study limiting cases of the model.
This includes the case of strictly 
zero anisotropy with and without biquadratic exchange,
and the case of infinite anisotropy with 
fixed four-fold orientation of the magnetizations
in each layer.
For the antiferromagnetic two-layer systems ($N$=2),
i.e. the sandwich structures (experimentally
realized as ferromagnetic/spacer/ferromagnetic trilayers),
we provide a detailed investigation of the
magnetic phase-diagram for arbitrary orientation
of fields in the layer planes.
%
%
Based on this, we can understand 
the \textit{basic} magnetic configurations 
in the multilayers, and we can give 
a map of the \textit{topologically different types} 
of magnetic phase diagrams.

We use standard methods to analyse magnetic
phases and transitions within the phenomenological approach
and the theory of phase transitions.\cite{Landau5} 
The one-dimensional chain models 
are considered as composite order-parameters
with many components 
($N$ components in a multilayer 
stack composed of $N$ ferromagnetic/spacer bilayers) 
and a characteristic structure of couplings 
between the components 
defined by interlayer exchange and 
the surfaces. 
From this point of view, 
the very rich phase-diagrams and correspondingly 
complex sequences of magnetic configurations 
can be understood.
For the general cases of the model, 
the equations for equilibrium and phase stability 
can be solved only by numerical methods.
With the methods and results expounded below, 
one can extend the analysis to specific
experimental cases in all detail. 

The solutions for the one-dimensional chain models
include various field-induced canted and inhomogeneous 
states with a net magnetization.
Based on the phase diagrams of these models, 
the evolution of laterally inhomogeneous 
(domain) states and magnetization processes
can be discussed.\cite{UFN88,FTT90} 
In this connection, the coexistence regions of
different phases in the vicinity of discontinuous
(first-order) magnetic phase transitions are important.
In external fields, 
thermodynamically stable domain-configuration 
from these competing phases can be established 
in extended multilayers 
This is crucial for an understanding of 
the hysteretic magnetization processes 
under coercivity mechanisms.

The paper is organized as follows.
After introducing the model and mathematical
tools (Sec. II) we consider the effects
of the bilinear and biquadratic
exchange interactions in the next
section (Sec. III). 
In Sec. IV we investigate in detail
four-fold anisotropy effects 
in antiferromagnetically coupled two-layers 
and then discuss the generalization
of  these finding to the case of multilayers.
In Sec. IV we discuss domain states and magnetization 
processes by using qualitative arguments.
In the concluding part we discuss possible 
extensions of the theory, and we suggest 
some useful experiments to enhance 
our understanding of magnetization 
processes in antiferromagnetic superlattices.

\section{The micromagnetic energy}
Let us consider a stack of $N$ ferromagnetic plates 
infinite in $x$- and $y$-directions and with finite thickness 
along $z$-axis. 
The magnetization of the layers is $\mathbf{M}_i$, 
and there are indirect interlayer-exchange couplings
through spaces between them. 
The phenomenological energy of the system can be written
in the following form 

\begin{eqnarray}
\Phi_N=\sum_{i=1}^{N-1} \left[ J_i\,\mathbf{m}_i \cdot
\mathbf{m}_{i+1}
+ \widetilde{J}_i\,\left(\mathbf{m}_i \cdot
\mathbf{m}_{i+1} \right)^2 \right]
-\mathbf{H}\cdot \sum_{i=1}^{N}\zeta_i \mathbf{m}_i
- \frac{1}{2}\sum_{i=1}^{N} K_i\left( \mathbf{m}_x^4 
+\mathbf{m}_y^4 \right) 
\label{energy1}
\end{eqnarray}
where ${\mathbf m}_i={\mathbf M}_i/M_0^{(i)}$
($M_0^{(i)}= |{\mathbf M}_i|$)
are unity vectors along the $i$-th layer  magnetization.
$\zeta_i = M_0^{(i)}/M_0$  designate deviations 
of the magnetization in the $i$-th layer from the
averaged value $M_0$.
We assume that  ${\mathbf m}_{iz}$=0 ,
i.e. the layer magnetizations are restricted to the layer plane. 
$J_i$ and $\tilde{J}_i$ are constants of bilinear 
and biquadratic exchange interactions, respectively;
$K_i$ are constants of the in-plane four-fold anisotropy.   
The functional (\ref{energy1}) generalizes similar models
considered earlier in a number of studies
on exchange \cite{Hubert98, Demokritov98, Chesman98, Ustinov01}
and anisotropy \cite{FTT90, Dieny90, Zabel94, Ustinov01}
effects in planar antiferromagnetic systems.
Within this approach the ferromagnetic 
layers are considered as homogeneously
magnetized blocks with 
constant values of the magnetic interactions.
This assumption deserves some comment.
It is well-established that in
magnetic nanostructures surface/interface
exchange and relativistic interactions
strongly modify electronic and magnetic 
properties within all volume of the 
magnetic constituents \cite{Kim01}.
This means that the values of the exchange or
anisotropy parameters, and the magnetizations 
include large interface/surface-induced components 
which may strongly
vary across the thickness of the individual layers.\cite{Kim01,PRL01}
However, the hypothesis
of magnetic homogeneity in the ferromagnetic
nanolayers in the models of type (\ref{energy1})
has a solid physical basis
and is justified by successful applications 
of these models to describe magnetization processes
in layered  ferro- and antiferromagnetic
nanostructures.\cite{Johnson96, Thiaville92, PRL01,
Wang94, Zabel94, Ustinov01, Lauter02, PRB03}
This relies on the fact
that in ferromagnetic nanolayers
the intrinsic (direct) exchange coupling 
are usually very strong and 
overcome surface/interface induced interactions.
Thus, they play the dominating role 
to determine the magnetic order 
\textit{within} the layers, which reacts 
also very stiffly on all external and induced
magnetic forces.
Furthermore, in these planar systems
the stray field forces confine the 
magnetization of the layers into their plane.
As a result, in most of these systems 
the ferromagnetic layers preserve 
essentially homogeneous in-plane magnetized states 
even under the influence of strongly inhomogeneous 
induced interactions.\cite{Thiaville92, PRL01}
In this connection it is important to
stress here that the
materials parameters in the phenomenological
model (\ref{energy1}) should be treated as
averages over a multilayer \textit{period}.
They comprise (in integral form) all 
intrinsic and induced energy contributions 
of the magnetic states in the ferromagnetic layers. 
In particular, for the systems
under consideration they may also
include a contribution from magnetism of 
the spacer layers.
Thus, in contrast to their bulk counterparts
which are considered as constants of
the magnetic material,
in nanoscale multilayer systems 
the phenomenological parameters 
strongly depend on many physical and geometrical
factors and may considerably vary from sample to sample.

Parity of $N$ also plays an important role.
The model (\ref{energy1}) 
with \textit{even} number of ferromagnetic layers
represents systems with fully compensated magnetization. 
Such superlattices can be treated as analogues 
to bulk collinear antiferromagnets.
%
%
Superlattices with odd numbers of layers 
or with unequal thicknesses of layers 
own a noncompensated magnetization 
which strongly determines
their global magnetic properties. 
In their response to an applied field,
these structures are similar to bulk 
ferrimagnets. 
They could be studied by similar methods as used below,
but have to be considered as separate class of systems.
Concentrating on the properties of systems with
fully compensated total magnetization,
we consider only superlattices with even $N$
and equal magnetization in all layers ($\zeta_i=1$).
%
%
In this contribution dedicated to general 
properties of these antiferromagnetic superlattices,
we avoid secondary effects which are related to 
their micro-structure and interface properties,
such as strains, chemical intermixing etc.
With that in view we will study the model (\ref{energy1}) 
mainly for the case of \textit{identical} 
ferromagnetic layers assuming that the exchange 
and anisotropy constants are equal for all 
the layers,
$J_i = J$, $\widetilde{J}_i = \widetilde{J}$, and $K_i = K$.
Some of the analytical results can be generalized 
to the isotropic (i. e. with $K_i=0$) 
system with  mirror symmetry about the center 
of the layer stack,
i.e.\ $J_i = J_{N-i}$ and 
$\widetilde{J}_i = \widetilde{J}_{N-i}$.

In our problem the magnetization of the \textit{i}'th layer
is confined in planar orientations
and can be described by the angle $\theta_i$ 
between the vector $\mathbf{m}_i$ 
and $x$-axis.
Thus, calculations of the magnetic states 
for the model (\ref{energy1})
can be reduced to the optimization
of the function $\Phi_N (\theta_1, \theta_2,...\theta_N)$.
Following the theory of  bulk antiferromagnetism
it is convenient to introduce here the vectors of the 
\textit{total magnetization} $\mathbf{m}$
and the \textit{staggered} 
(or \textit{antiferromagnetism}) vector 
$\mathbf{l}$
\begin{eqnarray}
\mathbf{m}=\sum_{i=1}^N \mathbf{m}_i, \qquad 
\mathbf{l}=\sum_{i=1}^N (-1)^{(i+1)}\mathbf{m}_i
\,.
\label{ml}
\end{eqnarray}
The energy  (\ref{energy1}) is invariant 
under transformation $\mathbf{l} \to -\mathbf{l}$,
and, therefore, all magnetic states 
in this model are degenerate with respect
to the sign of the staggered vectors.
In the following only solutions
with a definite sign of the staggered vector
will be discussed. However, one should keep 
in mind that the full set of solutions 
includes also those with opposite 
sign of $\mathbf{l}$.
The magnetic states with opposite sign of
$\mathbf{l}$ behave identically in an applied
magnetic field. Thus, for the magnetization 
processes there is no need to distinguish 
these solutions. 

For the particular case of a two-layer ($N=2$)
the energy (\ref{energy1}) can be transformed
to the following form
\begin{eqnarray}
\Phi_2= J_1\cos2\phi + \tilde{J}_1\cos^{2} 2\phi
 -2H\cos \phi \cos(\theta - \psi) 
-\frac{K}{4} \cos 4\phi \cos 4 \theta  - \frac{3K}{4}
\label{energy2}
\end{eqnarray}
where  $ \theta = (\theta_1+\theta_2)/2$, 
$ \phi = (\theta_1-\theta_2)/2$ and $\psi$ is the angle between
the  $\textbf{H}$ and the $x$-axis.

This case of a two-layer is of special interest.
The energy (\ref{energy2}) functionally coincides 
with that of a bulk two-sublattice antiferromagnet.
This offers the opportunity for
useful physical relations and analogies 
with bulk antiferromagnetism. 
On the other hand,
the two-layers represent the simplest 
model of antiferromagnetically coupled 
layered nanostructures. 

The equations minimizing 
the function $\Phi_N (\theta_1, \theta_2,...\theta_N)$
have strongly nonlinear character
and no analytical solution are available generally.
The body of our results have been obtained 
by numerical methods.
We solve the coupled equations
for equilibria $\{\partial W/\partial \theta_i =0\}_{i=1 \dots N}$
by an efficient conjugate gradient minimization.\cite{NR}
For systems with large anisotropy 
the configurations with the magnetizations oriented in the four-fold easy-axis directions 
are close to mostly metastable minima of the energy.
These configurations have been systematically 
searched for global absolute stability and checked for stability. 
Numerically, this is feasible for $N$ not too large 
with present day computers.
Stability limits and phase-transitions
are determined from the evolution of 
the smallest eigenvalue $e_0(H,K)$ 
of the stability matrix 
$(\partial^2 W/\partial \theta_i \partial \theta_j), i,j=1 \dots N$
under changing parameters in (\ref{energy1}),
i.e. ratios $\widetilde{J}/J$, $K/J$,
or the applied magnetic field.
\begin{figure}
\includegraphics[width=8.5cm]{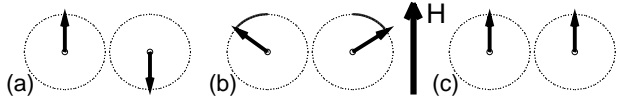}
\caption{
\label{BasicSpinsConfigs}
Basic spin-configurations in isotropic 
antiferromagnetically coupled two-layers ($N$=2). 
At zero field 
(a) the magnetization vectors $\mathbf{m}_i$  
are antiparallel in adjacent layers
but the structure has no fixed orientation
in the plane (infinite degeneracy).
An in-plane magnetic field lifts the degeneracy
and stabilizes states 
with \textbf{l} perpendicular to the field 
and with \textbf{m} along the field - 
this is a spin-flop (SF) phase (b).
In a sufficiently strong magnetic field the SF phases 
transform into the saturated (\textit{flip}) phase (c)}
\end{figure}

\section{Magnetic structures in isotropic multilayers.
Spatially inhomogeneous spin-flop states }

\subsection{Exchange interactions in layered antiferromagnets}
We start the investigations of the static configurations
minimizing energy (\ref{energy1})
from the isotropic case (i.e. with $K_i$ = 0).
Depending on relations between bilinear and
biquadratic constants different collinear and
noncollinear configurations are stable.\cite{Hubert98, Demokritov98})
The equations $J_i > 0$, $\widetilde{J}_i>0$,
$J_i-2\widetilde{J}_i >0$
determine the region in the parameter space where
a collinear 
\textit{antiferromagnetic} (AF)
phase is  the zero-field ground-state 
consisting of blocks of adjacent layers 
with antiparallel orientation of the magnetization   (Fig. 1(a)).
We restrict  our analysis to this 
practically important case. 
The AF phase is infinitely
degenerate with respect to orientations of 
the staggered vector \textbf{l} in planar directions.

The magnetic states for the isotropic two-layer ($N=2$)
in the applied field can be obtained by minimization of the energy (\ref{energy2})
with $K=0$. 
The solutions ($\theta = \psi$, 
$H= 2(J-2\widetilde{J})\cos \phi +8\widetilde{J} \cos^3 \phi$)
describe the states 
with the staggered vector perpendicular to the applied field 
(so called \textit{spin-flop} (SF) phase). 
In an increasing magnetic field
the magnetization vectors rotate 
into the field direction (Fig. 1 (b)),
and finally a continuous transition 
into the ferromagnetic (\textit{flip}) phase with $\tilde{\theta}=0$.
This (\textit{spin-flip} transition) occurs  at 
the exchange-field $H_e^{(2)} = 2(J+2\widetilde{J})$ (Fig. 1 (c)),
which is lower than the corresponding spin-flip field $H_e=4(J+2\widetilde{J})$
for a bulk system.

In superlattices with $N \ge 4$  the magnetic field applied 
in the plane also lifts 
the degeneracy and stabilizes the SF phase
(the state with  $\mathbf{l} \perp \mathbf{H}$ and
$\mathbf{l} || \mathbf{H}$) (Figs. 2 to 4).
However, the magnetic configurations in these SF states 
and their evolution in the applied field are markedly 
different from those in bulk antiferromagnets or 
in the two-layer systems.
It turns out that magnetic structures of the multilayers
with $N$ divisible 
by four ($N=4\,l$ called here \textit{even-even}) 
differ from \textit{even-odd} systems 
with $N=4\,l+2$ ($l= 1,2...$). 
In low fields ( ${J_i-2\widetilde{J}_i \gg  H}$)
the solutions for the SF phase
are given by a set of linear equations
\begin{eqnarray}
(J_{2j-1}-2\widetilde{J}_{2j-1})(\pi -\theta_{2j-1}+\theta_{2j}) = H,
\quad 
\theta_{2j} - \theta_{2j+1} =0, \quad j=1,2...l, 
\label{sol1}
\end{eqnarray}
where $l=N/4$ for even-even and
$l=(N+2)/4$ even-odd systems.
It is clear that for the isotropic model the direction of
the magnetic field in the film plane plays no role.
Following the definition of the angles $\theta_i$ we assume
here that the magnetic field is applied along the $x$-axis.
The solutions (\ref{sol1}) describe small  
deviations of the magnetization vectors,
from the directions 
perpendicular to the easy axis (see the magnetization
profiles in Figs.~2, 4 and the configuration in Fig. 3 (b)). 
In all internal layers ($i=2....N-1$) 
neighbouring pairs retain antiparallel orientations 
(e.g., ($\textbf{m}_2,\textbf{m}_3$) and ($\textbf{m}_4,\textbf{m}_5$) 
in Fig. 3b). 
For even-even systems the magnetization vectors of 
the central layers ($\textbf{m}_{N/2}$, $\textbf{m}_{N/2+1}$) 
preserve perpendicular orientation even in finite (but weak) fields 
(see layers 6 and 7 in Fig. 3). 
For even-odd systems the magnetization vectors of all layers 
have finite deviations from perpendicular orientation.
Towards top and bottom layer $i=1$ or $N$ in the stack, 
these deviations increase. 
Due to antiparallel
orientation the pairs have zero net magnetization. The total
magnetization arises only due to the top and bottom layers $i=1$ and $N$.
Note that in these configurations a number of layers 
rotates against the applied field 
(in Fig. 2-4 the layers with $i$ = 2, 4 and
their symmetric counterparts with $i$ = 11, 9).
This occurs because,
in weak fields, 
the exchange interactions 
favouring antiparallel magnetizations
in adjacent layers play the dominant role.
An increasing magnetic field counteracts 
and slows down this reverse rotation,
when the non-linear evolution for the magnetization
structure is reached,
and finally the sense of rotation is changed
at characteristic fields $H_r^{(i)}$ 
where $d\theta_i/dH=0$.
With increasing $N$  the number of layers 
increases which display this reverse rotation, 
the deviations from orientations perpendicular
to the applied low field
become larger near the surface layers,
and the fields $H_r^{(i)}$ are reached 
at lower fields (see inset in Fig. 2 for $N=64$).
A further set of characteristic fields $H_{\perp}^{(i)}$ defines 
the points where the projection of $\textbf{m}_i$ 
on the field directions changes the sign
(the field  $H_r^{(2)}$ and $H_{\perp}^{(2)}$ are indicated in Fig. 2.) 
In increasing field these
characteristic fields initially are reached 
for the central layers and at higher
fields for those closer to the boundaries. 
For fields $H > H_{\perp}^{(2)} = H_{\perp}^{(N-1)}$ 
the magnetization directions of all layers have
positive  projections onto the field direction. 
In the model with equal exchange
constant there is another special
field $H^*$ (independent of $N$),
where all inner layers have the same projection onto the
field direction ($\theta_i= (-1)^{i+1}\,\theta_0^{*}$, 
$i=2, 3...N-1$).
The parameters of this ``knot'' point are determined from 
a set of equations 
\begin{eqnarray}
H^*&=&4(J+2\widetilde{J})\left(1-\frac{2\,\kappa}
{1+\kappa}\sin^2 \theta^* \right)\cos \theta^*, \qquad \sin \theta^* = 2 \sin \theta^*_1,
\nonumber \\
&  &  \cos\left(\frac{\theta^*_1 + 3\theta^* }{2} \right) 
+ \kappa \cos\left(\frac{\theta^*_1 -\theta^* }{2} \right) 
\cos(\theta^*_1 + 3\theta^*)=0, \qquad  \kappa = 2\widetilde{J}/J.
\label{knot}
\end{eqnarray}
The functions $H^* (\kappa)$, $\theta^*_1 (\kappa)$, 
$\theta^* (\kappa)$ are plotted in Fig. \ref{KnotPoint}.
In particular for superlattices with zero
biquadratic exchange  $\tilde{J} = 0$
one has $H^*=\sqrt{6}\,H_e/4$ 
and 
$\theta^*=\mbox{acos}(\sqrt{3/8})$.
Near saturation, the SF phase  has only
positive projections of the magnetization 
on the direction of the magnetic field
which decreases towards the center 
similarly to spin configurations obtained numerically
in Ref.~\onlinecite{Nortemann92}. 
\begin{figure}
\includegraphics[width=8.5cm]{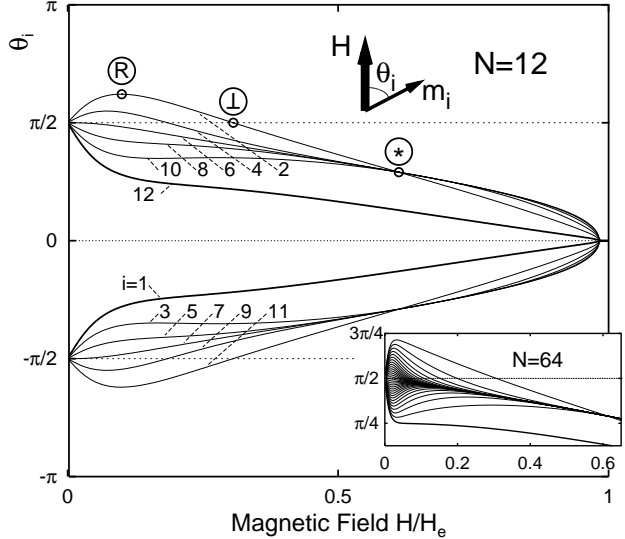}
\caption{
\label{EvolutionKzero}
Evolution of spin-states 
in multilayers
with a field $H$ applied 
in the plane of the layers and for zero anisotropy $K\equiv 0$ 
and biquadratic exchange $\tilde{J} = 0$, hence $H_e=4\,J$.
Main figure: rotation angles of the layers 
in the plane $\theta_i$ $i=1, \, \dots \,N$
for superlattice with $N=12$.
At point labeled ``R'' the sense of rotation
changes for the magnetization direction $\theta_i$ 
(here, $i=2$); at point ``$\perp$''  the magnetization
of this layer is perpendicular to the applied field again.
Point ``$\star$'' is 
the ``knot-point'' (Eq.~(\ref{knot})).
%
Inset: $N=64$ only $\theta_i$ for 
even $i=2$ and $i=4,8,12\dots 64$ shown.
}
\end{figure}
\begin{figure}
\includegraphics[width=8.5cm]{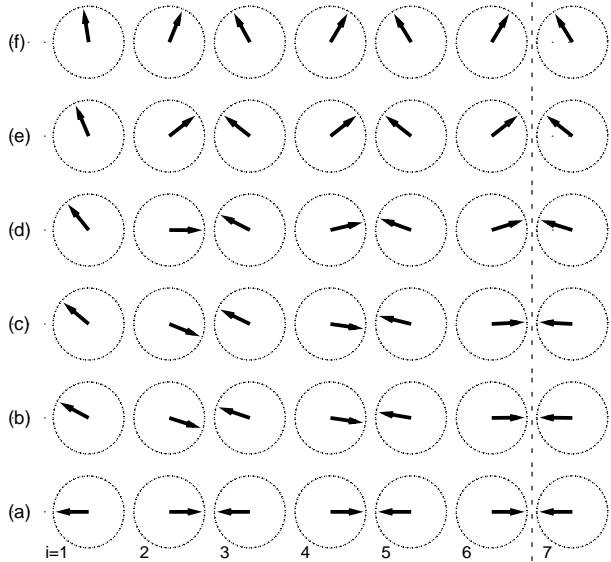}
\caption{
\label{StatesKzero}
Spin-states in multilayers with $N=12$
for zero anisotropy $K\equiv 0$
and field $H$ applied in the plane:
(a) $H=0$ (b) $H < H_R^{(2)} \ll H_e$
(c) $H=H_R^{(2)}$ (d) $H=H_{\perp}^{(2)}$
(e) $H=H^{\star}$, and (f) $H^{\star} < H < H_e^{(N)}$,
i.e. below the transition to the flip-phase
(cf. Fig. \ref{EvolutionKzero}).
All states have mirror-symmetry about 
the center of the multilayer stack 
(marked by the dashed line).
}
\end{figure}
The spatial inhomogeneity of the SF phases
in the multilayers and their remarkable
evolution in the applied field are due to
the particular \textit{finite-size}
effects in this type of magnetic nanostructures.
In such layered nanostructures
all internal layers interact with two
adjacent layers but 
the top and bottom layer
have only one neighbouring layer.
Thus, their exchange coupling 
is weakened compared to internal layers
within the multilayer stack.
This strong disbalance of the exchange forces 
at the boundaries affects the magnetic ordering 
within the entire multilayer 
and causes the reorientational processes 
in the SF phases.
This mechanism is also responsible for 
the occurrence of inhomogeneous states 
in other noncollinear structures
in the multilayers (see IV.C and V.A).
Note, this important role of the cut exchange-couplings
is specific 
to antiferromagnetic confined systems
with noncompensated moments at surfaces.
In contrast, the cutting of exchange couplings
does not affect \textit{ferromagnetically} 
coupled nanostructures where relative orientation
of the magnetization in the layers
essentially depend only on magnetic anisotropy, 
applied fields, and the demagnetizing stray-fields.

\subsection{Effects of the biquadratic exchange}
\begin{figure}
\includegraphics[width=8.5cm]{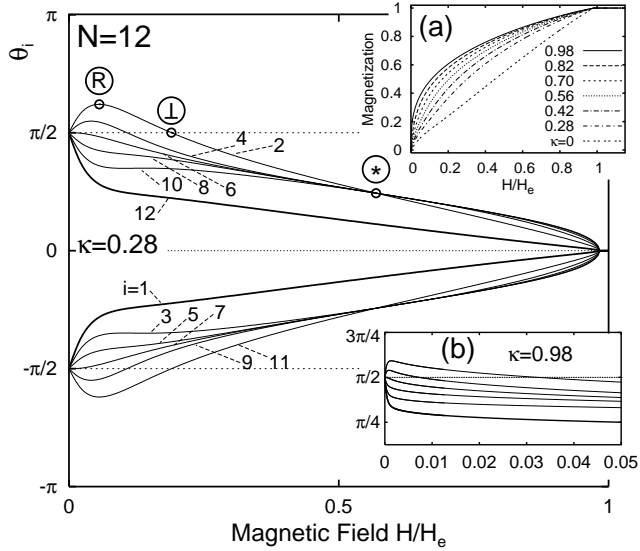}
\caption{
\label{EvolutionKappa}
Evolution of magnetic states 
in antiferromagnetic multilayer $N=12$
with zero anisotropy $K\equiv 0$
and biquadratic exchange $\kappa=2\widetilde{J}/J > 0$,
i.e. with enhanced  $H_e=4(J+2\widetilde{J})$
(compare with Fig.~\ref{EvolutionKzero} for $\kappa=0$).
$H$ is applied in the layer plane.
Insets (a): magnetization curves 
in the range of values of $\kappa$ 
with antiferromagnetic ground-state in zero field.
(b): evolution of states (only shown for 
layers with even numbers)
in small fields for $\kappa=0.98$.
}
\end{figure}

\begin{figure}
\includegraphics[width=8.5cm]{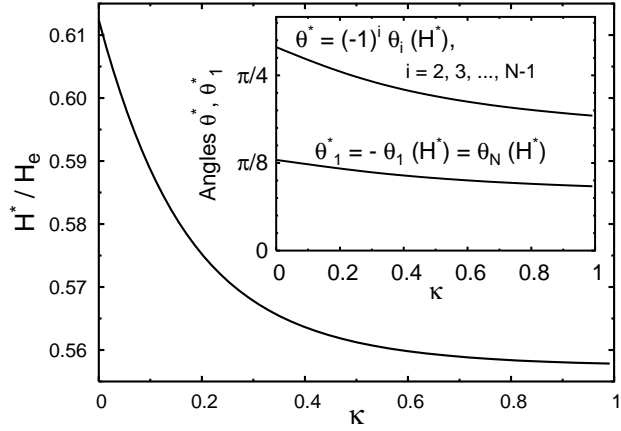}
\caption{
\label{KnotPoint}
Dependence of ``knot'' point
$H^{\star}$ (Eq. (\ref{knot})
on strength of biquadratic exchange $\kappa$.
Inset shows the two angles $\theta^*$, $\theta^*_1$
which characterise this special configuration
independent on $N$.
}
\end{figure}

Evolution of spin-flop profiles for the multilayer
with finite biquadratic exchange is plotted in Fig. 4.
For the systems with collinear ground-states
the biquadratic exchange does not induce
reorientional transitions but it has rather strong
influence on  the distribution of the magnetization 
in the multilayers (see insets in Fig.~4) and 
the value of the characteristic fields (Fig.~5).
The effects of a biquadratic exchange $\widetilde{J}>0$,
if present in multilayers with collinear antiferromagnetic 
ground-state, can be easily understood.
For magnetic configurations close to the antiparallel 
orientation of neighbouring layers, it softens the linear
exchange forces; 
for nearly parallel orientation the system becomes
stiffer instead.
The evolution of the magnetic states in the external
field is distorted accordingly by the presence
of the biquadratic exchange $\widetilde{J}>0$ 
(Fig.~\ref{EvolutionKappa}). 
In low fields close to the antiferromagnetic ground-state,
the system reacts more strongly on the external field.
Thus, the fields 
$H_{r}^{(i)}$,
where the rotation of the i'the layer reverses,
and the field $H_{\perp}^{(i)}$,
where perpendicular 
orientation with respect to field is regained,
respectively, 
are reached at relatively lower fields
(labels ``R'' and ``$\perp$'' in Fig.~\ref{EvolutionKappa}).
Conversely, the ferromagnetic state is reached only 
close to the enhanced exchange fields $H_e$
reflecting the stiffening of the effective exchange 
couplings in nearly ferromagnetic configurations. 
Thus, the magnetization curves acquire 
a nonlinear character with increasing $\widetilde{J}>0$ 
(inset (a) in Fig.~\ref{EvolutionKappa}).
Experimental observation of the fields $H_{r}^{(i)}$ 
and the special field $H^{*}$  (Fig. \ref{KnotPoint})
could be used to determine
the relative strength of $\widetilde{J}>0$. 
For large values of  $\widetilde{J}$ the reversal 
of the rotation of the magnetizations of individual
layers takes place at very low fields 
(see, inset (b) in Fig.~\ref{EvolutionKappa}).

\subsection{Spin-flip transition}
\begin{figure}
\includegraphics[width=8.5cm]{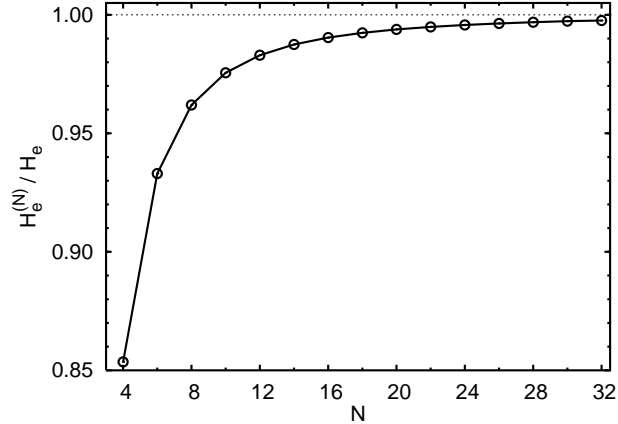}
\caption{
\label{HNe}
Spin-flip fields for antiferromagnetic multilayers
(zero anisotropy $K\equiv 0$) in dependence on
number of layers $N$.
}
\end{figure}
Near the spin-flip transition from the SF-phase to the
ferromagnetic states, the deviations of $\mathbf{m}_i$ 
from the field directions are small ($\theta_i \ll 1$), 
and the energy of the system (\ref{energy1}) 
can be expressed by the quadratic form
$W=\sum_{i,j=1}^{N} A_{ij}\theta_i\theta_j$, 
where the matrix  $A_{ij}$ 
has a tridiagonal band structure
with nonzero elements only in the main diagonal 
$A_{ii}=H-\bar{J}_i$ and with the side diagonal elements
$A_{i,i-1}=A_{i-1,i}=J_i+2\widetilde{J}_i \equiv \bar{J}_i$.
The spin-flip field $H_e^{(N)}$
is determined as the largest root of the equation 
$\mathrm{det}(A_{ij})=0$. 
In particular for the model with equal parameters 
these solutions are plotted in Fig.~6.
The spin-flip field gradually increases 
from $H_e^{(2)}= 2(J+2\widetilde{J})$ 
for the two-layer to the "bulk" value 
$H_e= 4(J+2\widetilde{J})$ in the
limit of infinite $N$ (Fig. 6). 
This dependence reflects the increase
of the average number of exchange bonds from the value 1 
for the two-layer and approaching 2 as $N$ tends to infinity.

\section{Effects of tetragonal anisotropy in two-layer systems}
Effects of the four-fold magnetic anisotropy 
on the states in the antiferromagnetic superlattices 
are revealed in detail by applying fields 
in arbitrary directions.
We approach the general case by a detailed
investigation on the case of the two-layer.
To set the stage, we discuss the highly symmetric
phase-diagrams with fields in direction of easy
and hard-axes in the plane. 
This completes earlier
work by Dieny et al. \cite{Dieny90}
on the $N$=2 systems.
Then, we present phase diagrams
for arbitrary field directions in the layer plane.
%
%
%
As we have seen in the last section, 
the primary effect of biquadratic exchange 
is a distortion of phase diagrams 
in various regions.
Thus, here we consider only models 
with $\tilde{J}=0$ to avoid 
such quantitative effects 
which will not affect 
the general topological features of the phase diagrams.

Optimization of the function $\Phi_2(\theta, \phi)$ 
Eq.~(\ref{energy2})
yields solutions for the magnetic states and 
their stability limits.
In zero field the tetragonal anisotropy lifts the 
rotational degeneracy of the AF phase
and stabilizes two different states with perpendicular
orientations of $\mathbf{l}$ (Fig. 7 (a)).
%
\begin{figure}
\includegraphics[width=7.6cm]{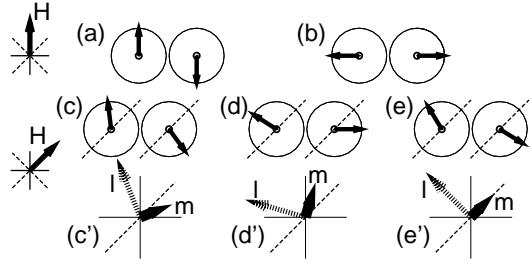}
\caption{
\label{TetragonConfigs}
Magnetic configurations 
in planar antiferromagnetically 
coupled two-layers with four-fold anisotropy. 
Easy directions are indicated by thin continuous lines,
hard directions by dashed lines.
In zero field 
two antiferromagnetic configurations (I) (a) and (II) (b) 
with the staggered vectors along the 
easy directions correspond to the ground-state.
A magnetic field along the hard directions 
distorts them into the canted phases (c), (d)
which transform into the spin-flop state (e).
Corresponding configurations of
net magnetization and staggered vectors 
are shown for the two canted phases (c') (d'),
and for the spin-flop phase (e').
}
\end{figure}
\begin{figure}
\includegraphics[width=7.6cm]{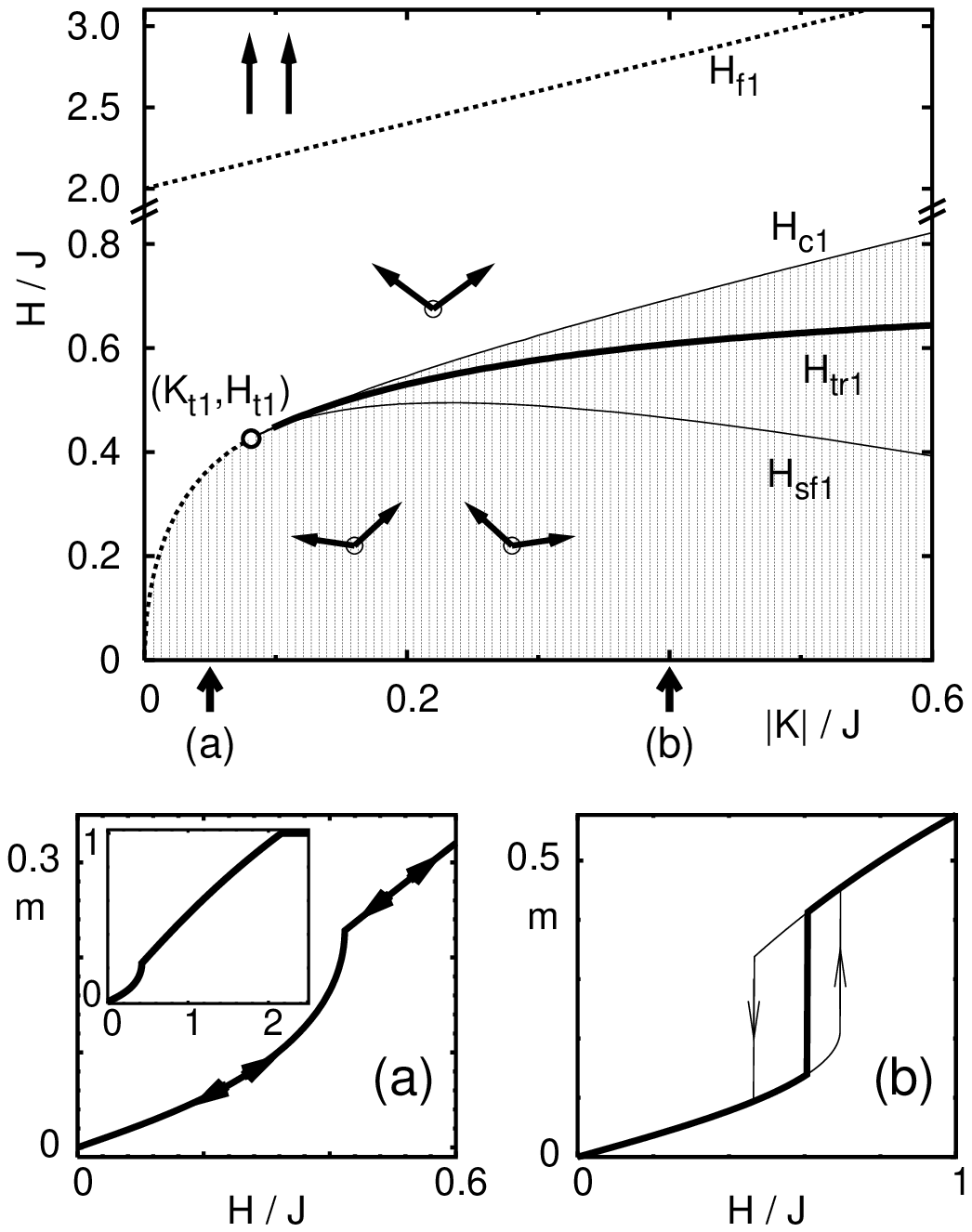}
\caption{
\label{KHhard}
Magnetic phase diagram of the two-layer system
in magnetic fields along \textit{hard} directions. 
The tricritical
point ($K_{t1}=0.081406\,J$, $H_{t1}=0.425780\,J$) 
separates the continuous (dotted line) 
and discontinuous (thick line) transitions between 
the canted and SF phases. 
A further dotted line gives
the critical field of the second-order
transition between  the SF and flip phases $H_{f1}$ 
(\ref{tran2}). 
Thin lines give the lower stability limit $H_{sf1}$ of the SF phases
and the upper limit $H_{c1}$ of the  canted phases. 
The grey  area is the region of the phase equilibrium 
between the two canted phases (see Fig.~\ref{TetragonConfigs} 
(c),(c') and (d),(d')).
Magnetization curves for low and high anisotropy,
as indicated in the main panel,
are plotted in (a) and (b).
}
\end{figure}

\subsection{ Magnetic field along hard directions}
In a field applied along one of the hard axes, 
the AF configurations of the ground-state
(Fig. 7(a),(b)) are distorted into low symmetry
configurations (so called \textit{canted} phases - Fig. 7(c),(d)). 
Both these magnetic configurations preserve mirror symmetry
with respect to the field direction and have the same energy.
The total magnetizations \textbf{m} of these canted states 
deviate from the field directions to different sides at equal
angles (Fig. 7(c'),(d')). 
An oblique magnetic field deviating from the hard direction
to one or the other side violates the phase balance between 
these two canted states. 
It favours the canted phase with the larger projection 
of \textbf{m} onto the field direction.
This is the typical situation of 
a first-order (or discontinuous) transition. 

In an increasing field 
the vectors \textbf{m} in both canted phases
rotate into the field direction and reach it
at a certain critical field ${H}_{sf1}$. 
At this point both phases transform 
into the common configuration corresponding 
to the SF phase (Fig. 7(d)), 
i.e. a second-order (or continuous) phase transition 
from low symmetry canted phases into the high-symmetry 
SF-phase occurs in this field. 
Standard analysis yields the following 
expression for the  parameters of the critical point
\begin{eqnarray}
H_{sf1} = (J+K\nu)\sqrt{2(1+\nu)},\quad
 \nu=(1+k-\sqrt{1+14k+25k^2})/(6k), 
 \quad k = |K|/J \,,
 \label{tran2}
\end{eqnarray}
where $\nu =\cos \phi_{sf}$  determines the equilibrium 
SF configuration at the critical field.

The two canted phases are competing phases related 
by a first-order transition. At the field $H_{sf1}$ 
the discontinuity between these phases disappears,
i.e. this is a \textit{critical end-point} 
of the first-order transition.
After the transition into the SF phase,
the system further evolves by rotation
of the magnetizations $\mathbf{m}_i$ 
into the direction of the field. 
This process is terminated at another
critical field $H_{f1}=2J(1+k)$ (Fig.~8).

The transition 
between the canted and SF phases is 
continuous only below a certain anisotropy strength, 
$|K| < K_{t1}$,
and becomes discontinuous at higher anisotropy.
The parameters of the corresponding tricritical point 
$K_{t1}=0.081406\,J$, $H_{t1}=0.425780\,J$
have been calculated numerically 
from the equation $K_{t1}=(1+\nu)/(6\nu^2-2\nu-4)$ 
together with Eqs.~(\ref{tran2}).
Numerically calculated transition fields~$H_{tr1}$
and the upper stability fields of the canted phases 
$H_{c1}$ are plotted in Fig.~8. 
In this picture the grey area 
describes the region of the 
first-order transition, respectively the 
region of phase-coexistence between the canted phases. 

\subsection{ Magnetic field along easy directions}
A magnetic field applied in one of the easy directions 
violates the energy balance between the two AF states.
The state AF(I) with staggered vector parallel 
to the field does not change its configuration 
and exists as a metastable collinear state up 
to the critical field $H_{af}$.
The phase AF(II) with the staggered vector 
perpendicular to the magnetic field
transforms into a SF phase 
which corresponds to the global energy minimum of the system. 
In an increasing magnetic field this phase evolves similarly
to the SF phase along the hard directions, 
and continuously transforms
into the flip states at the critical fields $H_{f2}$ 
for low anisotropy.

The transition between the SF and flip phases changes 
the order in another tricritical point ($K_{t2},H_{t2}$).
At a critical field  $H_{sf2}$ (\ref{stab1})
the SF phase becomes unstable with respect to flipping distortions, 
i.e. modes that redirect the magnetization vectors
into the field direction but preserve 
the symmetry of the SF phase.
The parameters of the tricritical point ($|K|_{t2}=(1/5)\,J, H_{t2}=(8/5)\,J$).
and the stability limits 
\begin{eqnarray}
H_{af}= 2J\sqrt{k(1+k)},\quad 
H_{sf2}= \frac{2J}{3}\sqrt{\frac{2(1+k)^3}{3k}},\quad 
 H_{f2}= 2J(1-k)
\label{stab1}
\end{eqnarray}
have been calculated in \cite{Dieny90}. 
After some algebra the first-order transition field $H_{tr2}$
also can be derived in analytical form
\begin{eqnarray}
H_{tr2}=\frac{8J}{9}\left[1+\frac{(3+k)}{12}
\left(\sqrt{1+\frac{3}{k}}-1 \right) \right]\,,
\label{tran1}
\end{eqnarray}
At the transitional field we have $\cos \phi = (\sqrt{1+3/k}-1)/3$.
\begin{figure}
\includegraphics[width=7.6cm]{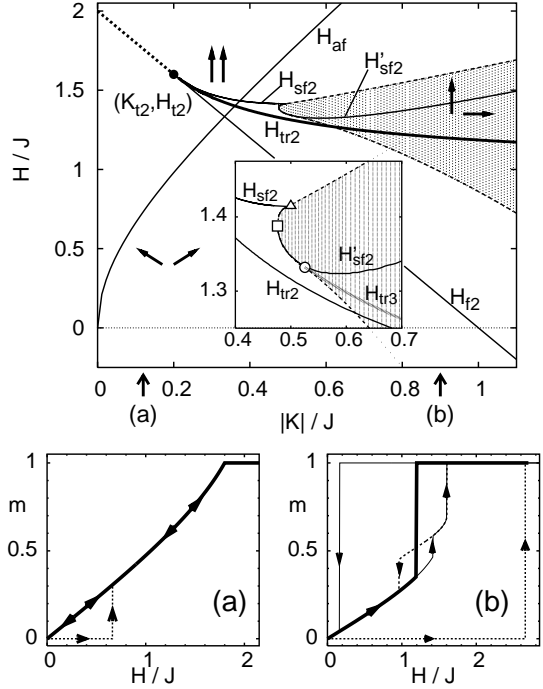}
\caption{
\label{KHeasy}
Magnetic phase diagram of the two-layer system 
in magnetic fields along \textit{easy} directions. 
The tricritical point ($K_{t2}=(1/5)\,J$, $H_{t2}=(8/5)\,J$) 
separates the continuous $H_{f2}$ (\ref{stab1}) (dashed line) 
and discontinuous $H_{f2}$ (\ref{tran1})(thick line)
transitions between SF and flip phases. 
The tricritical point ($K_{t2}=0.2J$, $H_{t2}=1.6J$) separates
the continuous $H_{f2}$ (\ref{stab1}) (dashed line) and discontinuous
$H_{f2}$ (\ref{tran1})(thick line)
transitions between SF and flip phases.
Thin lines indicate the lability lines of the individual
phases (\ref{stab1}).
The grey area designates the stability region
of the canted phases.
For the anisotropy $|K|/J=k^*$ = 0.47596 (point $\Box$),
the critical field of the SF phase
switches from an  instability against symmetric ``flipping''
at the line $H_{sf2}$ to an instability against canting 
at $H'_{sf2}$ given by the two branches 
$H^{(1,2)}_{sf2}$ from Eq.~(\ref{stabSF2}).
(Inset in main figure shows the details of this process; 
see text following Eq.~(\ref{stabSF2}) for further explanation.)
Small panels (a) and (b) show magnetization
curves for low and high anisotropies as indicated
in the main figure.
Full thick lines are for the evolution of the
equilibrium ground-states, in (a) from SF phase
to a continuous spin-flip transition into saturation.
Dotted lines are the evolution starting from the
metastable AF-phase. 
In (b) the evolution of the
saturated state in decreasing fields is given
by a thin line. The inner small hysteresis loop
shows the transitions from and into the metastable canted phase.
}
\end{figure}
The (K, H)-phase diagram for this cases
is plotted in Fig.~9. 

With increasing strength of the anisotropy,
the ``landscape'' described 
by the energy function $\Phi_2(\theta, \phi)$
acquires additional folds.
In particular, metastable canted
phases arise in a certain region
of the magnetic fields along the easy directions. (This
region is indicated by grey colour
in Fig. \ref{KHeasy}).
For $|K|/J \equiv k > k^*$ with $k^*$ = 0.28+0.08$\,\sqrt{6}$ = 0.47596  
the SF phase undergoes an instability with respect
to a transition into the canted phases.
This instability occurs at lability fields
which are given by the two branches of the following parametric
equation
\begin{eqnarray}
H_{sf2}^{(1,2)} = (J-|K|\nu_{1,2})\,\sqrt{2(1+\nu_{1,2})},\quad
 \nu_{1,2}=(k-1 \mp \sqrt{1-14k+25k^2})/(6k), 
\label{stabSF2}
\end{eqnarray}
where $\nu_{1,2}$ is given by the the configuration 
in the SF-state, $\nu_{1,2} = \cos2\phi_{sf2}$.
Both branches of the lability field (\ref{stabSF2}) start
at the point $H_{sf2}^{(1,2)} (k^*)= 1.3895\,J$
(point $\Box$ in Fig. \ref{KHeasy});  
$H_{sf2}^{(2)}$ meets the lability line $H_{sf2}$ of Eq.~(\ref{stab1})
at the ``beak'' ($k = 1/2, H = \sqrt{2}\,J$) (point $\bigtriangleup$
in Fig. \ref{KHeasy}). 
The transition between the metastable SF phase
and the metastable canted phases
along  $H_{sf2}^{(1,2)}$ is continuous
between the points $\bigtriangleup$
and a further tricritical point $\bigcirc$.
For stronger anisotropy the transition between
these metastable phases is discontinuous
at a first-order transition field $H_{tr3}$.
This transition line along with the stability
limits of the canted phases have been
numerically determined as given in Fig.~\ref{KHeasy}.
Note that such processes 
between metastable phases
may be realized only in systems
with high coercivity where the transition
into the thermodynamically stable flip phase 
is suppressed at the field $H_{tr2}$ (\ref{tran1}).

\subsection{ Evolution of ($H_x,H_y$)-phase diagrams}
\begin{figure}
\includegraphics[width=7.6cm]{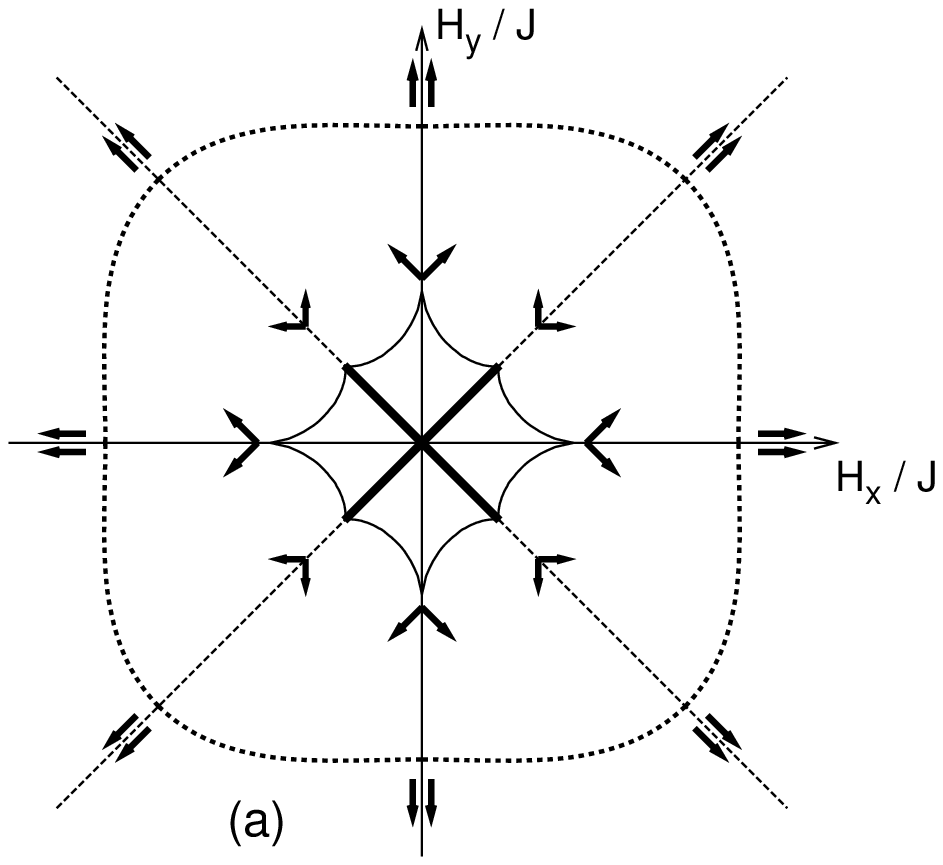}
\includegraphics[width=7.6cm]{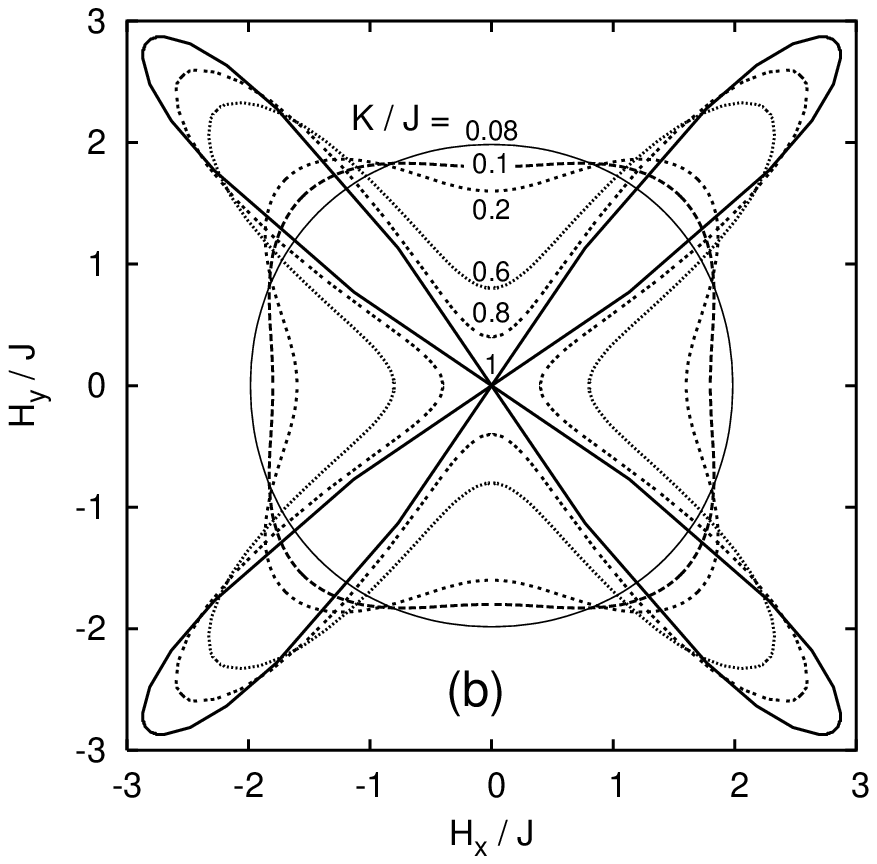}
\caption{
\label{HxHyweak}
(a) Schematic ($H_x,H_y$)-phase diagram of the two-layer
for $ 0 < K < K_{t1}$. The astroid (continuous line)
gives the stability limit of the canted metastable phases.
Thick lines in diagonal directions (hard axis) are 
first order transitions between canted phases.
The line for the (continuous) transition into 
the saturated spin-flip state is given by the
dotted line.
Note that the real size of the astroid 
in the ($H_x,H_y$)-diagram is much smaller
than shown here.
(b) Evolution of spin-flip field, respectively 
lower stability limits of the saturated
(ferromagnetic) phase (FM). 
Phase FM is stable outside the closed curves
for the various values of anisotropy $0<K<1$.  
For $K / J > 1.0$ the FM-phases are (meta)stable even 
in reverse fields 
and the existence regions overlap
for states saturated in different easy directions.
}
\end{figure}

Now we consider the magnetic states in applied fields 
deviating from the symmetric directions
and construct ($H_x,H_y$)-phase diagrams 
for different values of the parameter $k = |K|/J$.
For the ($H_x,H_y$)-diagrams in Figs.~\ref{HxHyweak} to 14 it is assumed
that the easy directions coincide with the $x$- and $y$-axis.
This corresponds $K>0$ in the energy Eq.~(\ref{energy2}). 

Diagrams for $K <0$
can be obtained by rotation of those for  
$K>0$ through an angle $\pi/4$.
In the limit of weak anisotropy, $|K| \ll J$,
independent minimization with respect to the angle $\phi$
yields $2J \cos \phi = H \cos(\theta - \psi)$
and the energy (\ref{energy2}) is simplified to 
the following form
\begin{eqnarray}
F(\theta) = - \frac{H^2}{2J}\cos^2 (\theta-\psi) -
\frac{K}{4J}\cos 4\theta\left[1+\frac{2H^2}{J}\cos^2 (\theta-\psi)\right]
\,.
\label{energy3}
\end{eqnarray}
The set of equations for the stability limits of 
solutions,  $dF(\theta)/d\theta = 0$ and
$d^2F(\theta)/d\theta^2 = 0$, yield
two closed lines of critical fields 
in the ($H_x, H_y$)-phase diagram (Fig.~\ref{HxHyweak}).
One of them $H= 2(J-K\cos 4 \psi)$
describes the second-order transition 
into the flip phase with $\phi = 0$.
The other closed curve can be written in
a parametric form
\begin{eqnarray}
H^2\cos (2\theta -2\psi )=-4KJ \cos 4\theta + K^2\Omega_1(\theta), 
\nonumber\\
H^2\sin (2\theta -2\psi )=-2KJ \sin 4\theta + K^2\Omega_2(\theta) \,,
\label{astroid1}
\end{eqnarray}
where $\Omega_1(\theta)= 4K^2( 7\cos ^24\theta
-2+2\text{sgn}(K)\sqrt{3\cos ^24\theta +1}$,
$\Omega_2(\theta)= 2K^2( 5\cos 4\theta
+2\text{sgn}(K)\sqrt{3\cos ^24\theta +1}$.
It describes an  astroid with eight-cusps 
that confines  the region of 
the canted metastable states (Fig.~\ref{HxHyweak}(a)).
\begin{figure}
\includegraphics[width=7.6cm]{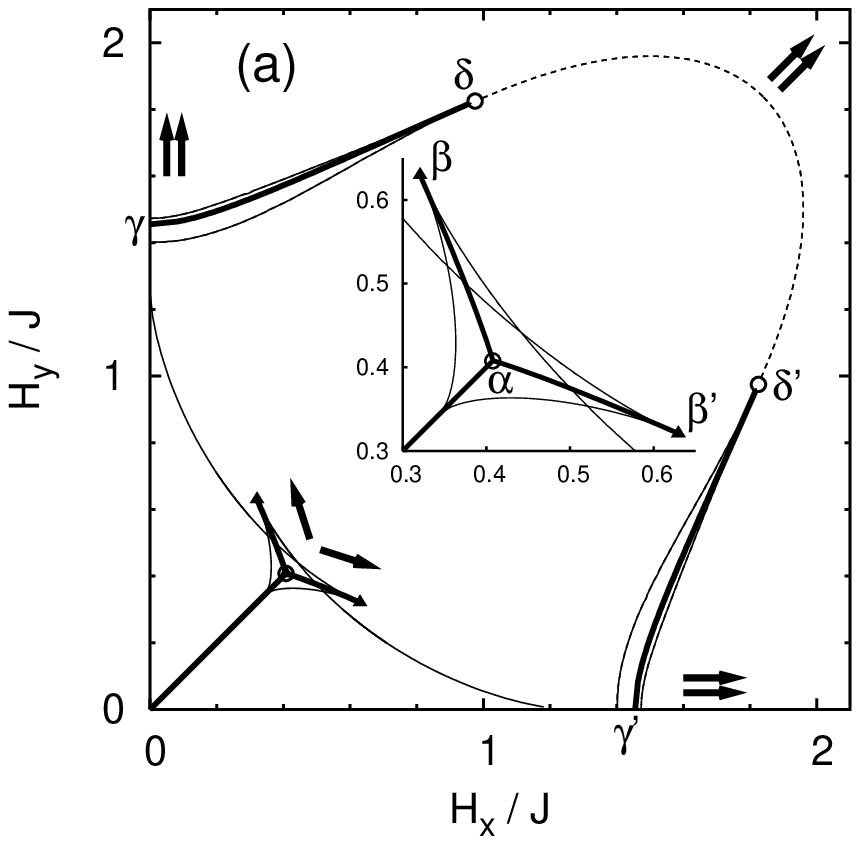}
\includegraphics[width=7.6cm]{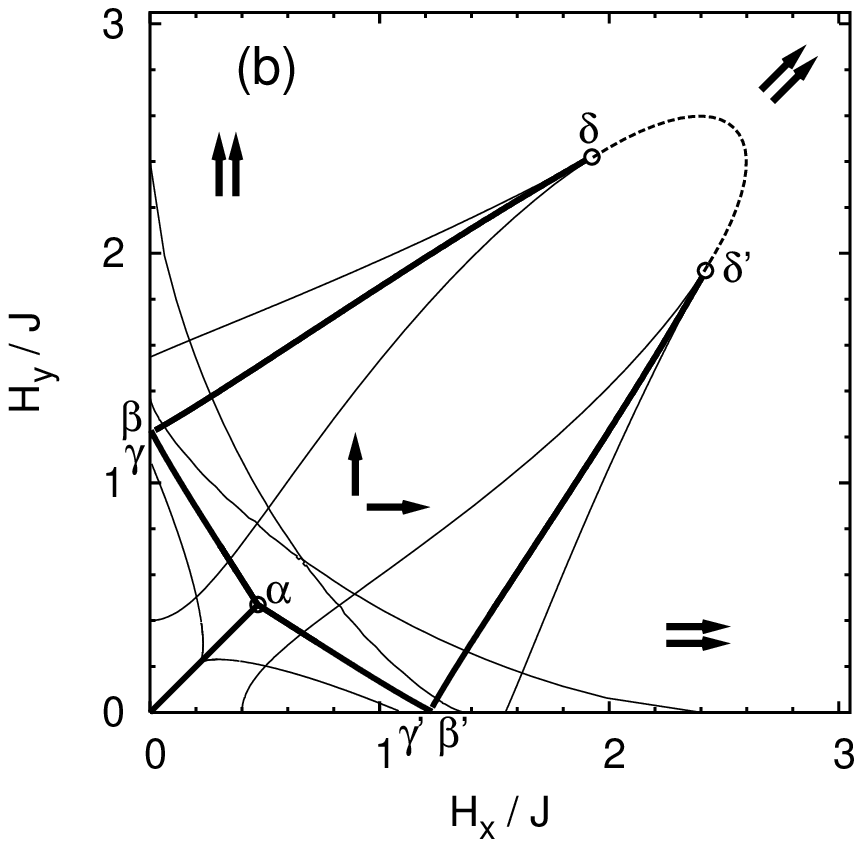}
\caption{
\label{HxHyO3}
First quadrant of the ($H_x,H_y$)-phase diagram for 
antiferromagnetic two-layers
with (a) $ K = 0.3\,J$ (b) $K = 0.8\,J$.
Easy axes of magnetic anisotropy are along the 
axes of the diagrams.
Inset in (a) magnifies the region of the ``swallow tail''
for the discontinuous transitions in magnetic
fields oriented close to the hard axes.
Thick lines are lines of first-order transitions.
The dotted lines $\delta$ -- $\delta'$ are
continuous spin-flips 
from spin-flop phase to the saturated (spin-flip) state.
Thinner lines are the limits of stability, 
defining the coexistence regions 
for the corresponding phases.
}
\end{figure}
The cusps along the easy directions coincide with 
the stability field of the AF phase $H_{af}$ Eq.~(\ref{stab1}),
and those along the hard directions with
$\tilde{H}_{sf1}$ Eq.~(\ref{tran2}). 
Thick black lines
within the astroid and along the hard directions
indicate the first-order transitions
between the canted phases from Fig.~\ref{TetragonConfigs}. 
This topology of the ($H_x,H_y$)-phase diagram Fig.~\ref{HxHyweak}(a)
is preserved  up to the first tricritical point  $K_{t1}$.
For $|K| \geq K_{t1}$
a first-order transition between 
SF phase and two canted phases 
arises along the hard
directions (point $\alpha$ in Fig.~\ref{HxHyO3}(a)).
For stronger anisotropies, $|K| > K_{t1}$,
discontinuous phase transitions
exist also for finite deviations of 
the field direction
from the hard axes. 
The corresponding line of these first-order transitions 
have critical end-points,
where the difference between the competing phases disappears
(analogously to the end-point of 
a coexistence-line
in gas-liquid phase diagrams).
For $|K| \geq K_{t2}$ 
and for fields along the easy directions,
the spin-flip into the 
saturated (induced) ferromagnetic state
occurs discontinuously. 
Thus, another line of first-order transitions
develops for $|K| > K_{t2}$ also
in oblique fields. 
Correspondingly, 
the transition lines for spin-flips in
($H_x,H_y$)-phase diagrams
consist of continuous and discontinuous
sections joined by the tricritical points $\delta$, $\delta'$.
The calculated ($H_x,H_y$)-phase diagrams 
for $K = 0.3\,J$ in Fig.~\ref{HxHyO3}~(a) 
and for $K=0.8\,J$ in Fig.~\ref{HxHyO3}~(b)
represent the main topological features 
of these phase diagrams.
Point `$\alpha$' indicates the field of phase equilibrium between
the SF phase and two canted phases and corresponds 
to the critical field $H_{tr1}$ from Fig. 8; 
Points $\beta$, $\beta'$ are the end-points of the first-order transitions
between one of the canted phases and the SF phase. 
Points $\gamma$ $\gamma'$ correspond to the 
transition field $H_{tr2}$ from Fig. 9;
points $\delta$, $\delta'$ are the tricritical points,
where the discontinuous transition 
between the flip phase and the distorted SF phase ends.
The lability lines 
in the vicinity of the transition
field $H_{tr1}$ have the shape of a ``swallow tail''. 
Similar phase diagrams arise in uniaxial ferromagnets 
with strong fourth-order anisotropy.\cite{Mitsek74, FTT87}
As the anisotropy constant increases in the region
$|K| > K_{t2}$ the lines of the first-order transitions extend 
and near the special value $|K|/J$=$k^{*}=0.476$ 
the points $\beta$, $\beta'$ reach the easy  directions. 
For larger $|K|$ canted phases 
exist as metastable state 
also for fields in easy directions (compare Fig.~9).
However, for increasing fields close to this direction, 
the discontinuous transition from the spin-flop state 
into the spin-flip state occurs before the transition 
into such a canted phase can take place, i.e.
the two first-order lines $\alpha$ -- $\beta$ 
and $\delta$ -- $\gamma$ cross each other close 
to the easy axis directions in Fig.~11(b).
Br{\"o}hl  et al. \cite{Zabel94} reported
evidence of an intermediate canted state
in a Co/Cu/Co/Cu(001) system with field 
in an easy direction, following the
theoretical prediction of such states
by Dieny et al. \cite{Dieny90}
However, the state was found 
at lower fields than expected. 
This may be due to a misorientation 
of the external field and/or a mosaic of the 
epitaxial layer system because then the 
canted state becomes stable already at lower fields,
as can be seen from line $\alpha$ -- $\beta$
in phase diagram Fig.~11(b).
Generally, it is not sufficient to 
investigate only magnetization behaviour 
in hard and easy axes directions for 
a thorough understanding of the magnetization
phases in these multilayer systems.

\begin{figure}
\includegraphics[width=8.5cm]{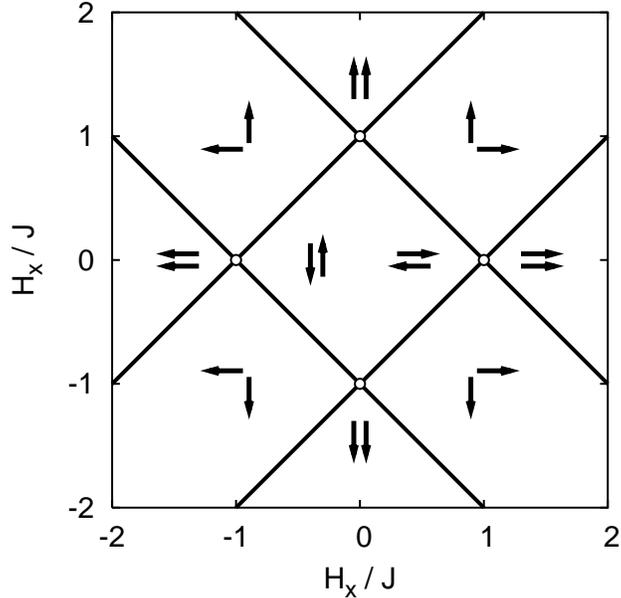}
\caption{
\label{HxHyinfinite}
($H_x,H_y$)-phase diagram of the $N=2$-layer system
in the limit of an infinite positive four-fold
anisotropy.
}
\end{figure}

To understand the transformation 
of ($H_x,H_y$)-diagrams with increasing tetragonal anisotropy
the limiting case is useful
where $\mathbf{m}_i$ are strictly 
oriented along the easy axes.
In this case with \textit{infinite} anisotropy, 
i.e. $|K|=\infty$,
our model can be considered as a chain of an antiferromagnetic
four-state clock-model or planar Potts-model\cite{Wu82}
in a transverse external field.
Recently, such four-state-clock models 
were employed to analyse the 
spin configurations of 
bulk tetragonal metamagnets 
with large four-fold anisotropies,
such as rare-earth nickel-borocarbides
\cite{Canfield97,Muller01}
or rare-earth silver-antimonides.\cite{Myers99}
Rich experimental ($H_x,H_y$)-phase diagrams 
have been obtained and analysed 
in terms of four-state-clock models,
e.g. for HoNi$_2$B$_2$C \cite{Canfield97} 
or DyAgSb$_2$.\cite{Myers99}
The values of 
the angles $\theta_i=$
0, $\pi/2$, $\pi$, or $3\,\pi/2$
in these four-fold states for $K=\infty$
are symbolically given by 
$\uparrow$, 
$\leftarrow$,
$\downarrow$, 
and 
$\rightarrow$, 
e.g. the AF and SF phases are
($\uparrow$\,$\downarrow$) and ($\rightarrow$\,$\leftarrow$) 
for $N=2$.
In addition to the collinear AF and 
flip states there are phases 
with perpendicular orientation of the magnetization,
or ``90~degree-folded''  phases
(Fig.~\ref{HxHyinfinite}).
The states created by all combinations of 
these four angles $\theta_i$ for the magnetization
in the multilayer stack
exist as metastable states in 
arbitrary magnetic fields
because the energy wells corresponding 
to these solutions are separated 
by infinitely high potential barriers.
The regions of the absolute stability of 
these phases are separated 
by first-order transitions lines 
shown in Fig.~\ref{HxHyinfinite}.
In the first quadrant the transition between
the AF states and 
the canted phase ($\uparrow\,\rightarrow$) 
occurs at the line $H_x+H_y=J$; 
the transitions lines 
from the canted phase ($\uparrow\,\rightarrow$)
into the flip states 
($\uparrow\,\uparrow$)
and  ($\rightarrow\,\rightarrow$)
are $H_y-H_x=J$ and  $H_y-H_x=-J$, correspondingly (Fig. 12).
In the points (1,0) and (0,1) four phases coexist.
Thus,  under increasing strength of the anisotropy
($K >0$) the ($H_x,H_y$)-phase diagram 
evolves from that plotted in Fig.~\ref{HxHyweak}(a) for $K \ll J$ 
to that in Fig.~\ref{HxHyinfinite} 
for infinitely large values of $K$.

\section{Magnetization configurations and processes in multilayers}
The equilibrium magnetic configurations 
in two-layer systems arise as a results of the competition 
between the interlayer exchange coupling and tetragonal anisotropy. 
For multilayers the disbalance of the exchange forces 
at the boundaries (Sec. II) additionally influences the magnetic states. 
We first describe the structure of the one-dimensional solutions 
for laterally homogeneous states in finite antiferromagnetic superlattices.
These magnetic states are determined by the interplay 
between cut exchange at the surfaces and the four-fold anisotropy.
Then we discuss some consequences for multidomain states
and magnetization processes, and we discuss the physical
nature of other effects which may play a role for the
magnetic behaviour of real experimental systems.

\subsection{Exchange cut versus tetragonal anisotropy}
First, we investigate
the ($H_x,H_y$)-diagrams 
of magnetic states of multilayers with $N \ge 4$
in the limit of infinite
four-fold anisotropy.
\begin{figure}
\includegraphics[width=7.6cm]{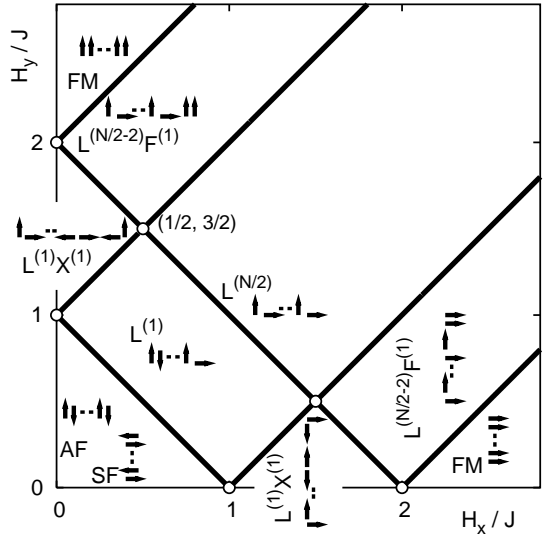}
\caption{
\label{HxHyinfinite2}
($H_x,H_y$)-phase diagram of the multilayer
in the limit of infinite positive four-fold anisotropy.
The phase-region of the L$^{(1)}$X$^{(1)}$-phases coincides
with that of the collinear ferrimagnetic F$^{(1)}$, 
not shown here. (For details see text.)
}
\end{figure}
The zero-field ground-states 
are the AF phases
$(\uparrow\,\downarrow\,\dots\,\uparrow\,\downarrow)$
and
$(\rightarrow\,\leftarrow\,\dots\,\rightarrow\,\leftarrow)$
%
We may restrict the field 
to be oriented in directions $\psi$ 
in the interval $[$0,$\pi$/4$]$.
The configurations for other values
of $\psi$ follow from symmetry. 
Then, we may distinguish the four ground-state (zero-field)
domains 
(AF1)$^{(n)}$= $(\uparrow\,\downarrow)^{(n)}$,
(AF2)$^{(n)}$= $(\downarrow\,\uparrow)^{(n)}$,
(SF1)$^{(n)}$= $(\rightarrow\,\leftarrow)^{(n)}$, and
(SF2)$^{(n)}$= $(\leftarrow\,\rightarrow)^{(n)}$,
where $n$ signifies the number of repetitions
of the pair in a domain.
In external fields $H>0$ configurations 
with net magnetization will be stabilized.  
These configurations must have pairs 
of adjacent moments flipped 
by 90 degree  L=$(\uparrow\,\rightarrow)$ 
or  X=$(\leftarrow\,\uparrow)$,
and by 180 degree F=$(\uparrow\,\uparrow)$.
The 90-degree-folded configuration ``X'' is 
less favourable than ``L'' for 
field orientations in the chosen range, 
except for fields in easy-axis direction $\psi=0$.
Thus, the magnetized configurations 
with lowest magnetization
and smallest expense of exchange energy 
are those with only one pair of type ``L''.
In a short-hand, we write L$^{(1)}$ 
for these configurations of type
(AF1)$^{(n)}\,/$L$\,/$(SF2)$^{(N/2-n-1)}$
(with $n=0,1,\dots$.)
Next, we may form configurations with 
higher magnetization and smallest expense 
of exchange energy
using one ``L'' and one ``X''-pair:
L$^{(1)}$X$^{(1)}$=
(AF1)$^{(n)}\,/$L$\,/$(SF2)$^{(N/2-n-m-2)}\,/$X$\,/$(AF1)$^{(m)}$,
($n,m=0,1,\dots$).
These configurations have the same energy and, therefore, are
degenerate with configurations containing only one ``F''-pair:
F$^{(1)}$=(AF1)$^{(n)}\,/$F$\,/$(AF2)$^{(N/2-n-1)}$.
Note that these states are highly degenerate because 
the ``L''-, ``X''-pair, or ``F''-pair
may be placed at arbitrary positions 
in the stack of $N$ layers.
%
%
In higher fields configurations with 
various combinations of ``L''-, ``X''-, and ``F''-pairs 
may be stabilized by an external field.
However, the structures with lowest energy,
i.e. the absolutely stable states, 
are rather simple because of the following considerations.
If, in external fields, 
more than one ``L''-pair can be formed starting
from the L$^{(1)}$ structure,
the formation of the maximum number 
of L-pairs gives the most favourable configuration. 
In particular,
the structure L$^{(N/2)}$=($\uparrow\,\rightarrow$)$^{(N/2)}$ 
is the lowest energy structure 
for fields pointing 
in hard axis-directions $\psi=\pi/4$ 
in the limit of infinite H.
The state with higher magnetization 
than L$^{(N/2)}$ for fields pointing closer 
to the easy axis direction, $0 \leq \psi < \pi/4$, 
and with the smallest expense of exchange energy
is in our notation L$^{(N/2-1)}$F.
The saturated state 
FM$\equiv$F$^{(N/2)}$ is the most favourable state 
which is reached whenever energy can be gained
in external fields by flipping more moments 
into position $\uparrow$
in the states L$^{(N/2-1)}$F, F$^{(1)}$, 
or L$^{(1)}$X$^{(1)}$.
Thus, for general $N$ we have only the following
phases for infinite positive four-fold anisotropy:
AF or SF as zero-field ground-states;
for fields with orientation close to the hard direction $\psi=\pi/4$ 
L$^{(1)}$ and L$^{(N/2)}$;
otherwise in intermediate fields 
two degenerate phases F$^{(1)}$ and L$^{(1)}$X$^{(1)}$;
at high fields an asymmetric L$^{(N/2-1)}$F-phase 
and the fully saturated ferromagnetic phase FM.
%
%
We did numerical checks to ascertain that 
no further energetically stable phases 
do exist in external fields of 
arbitrary strength and direction, indeed.
Thus, we searched for the states
of lowest energy by sampling all possible 
configurations for models with $N$=4, $\dots$, 12
corroborating our arguments.
Based on this set of magnetic
configurations the resulting ($H_x,H_y$)-phase diagram 
for general $N>4$ 
can be calculated analytically (Fig.~\ref{HxHyinfinite2}).
As in the simpler case 
of Fig.~\ref{HxHyinfinite} 
for the two-layer system
all these states are separated 
by infinitely high potential barriers 
and remain metastable for arbitrary fields. 
The first-order transitions
between different phases occur along straight lines 
as shown in Fig.~\ref{HxHyinfinite2}. 

\begin{figure}
\includegraphics[width=7.6cm]{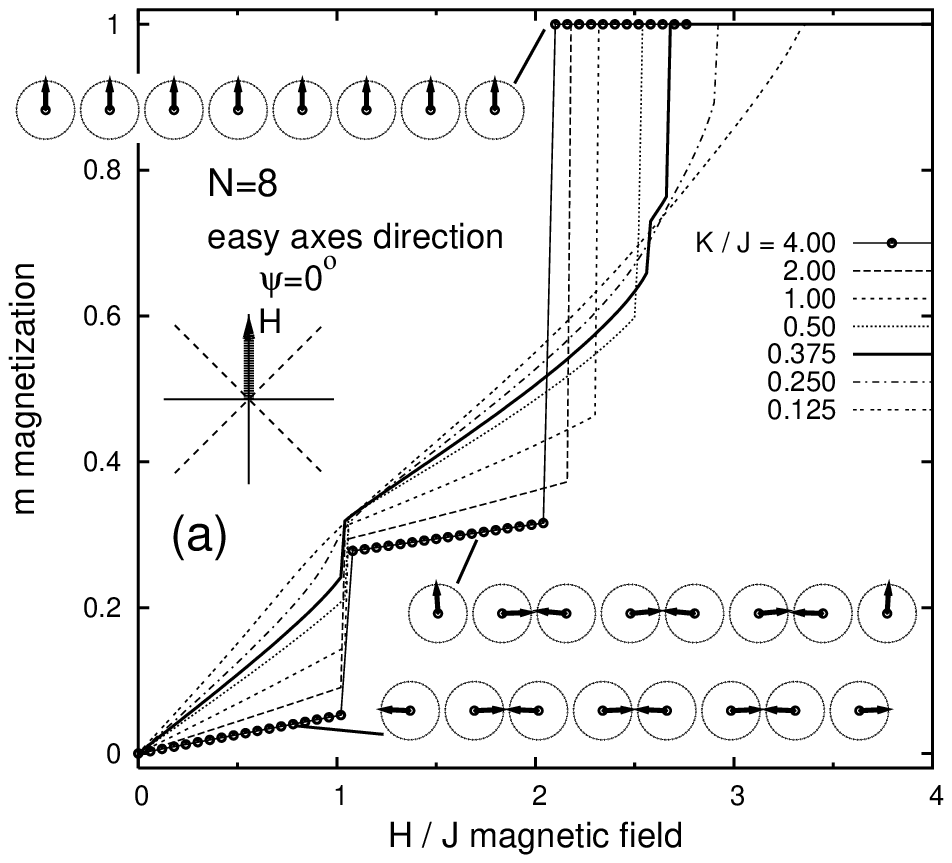}
\vspace{12pt}
\includegraphics[width=7.6cm]{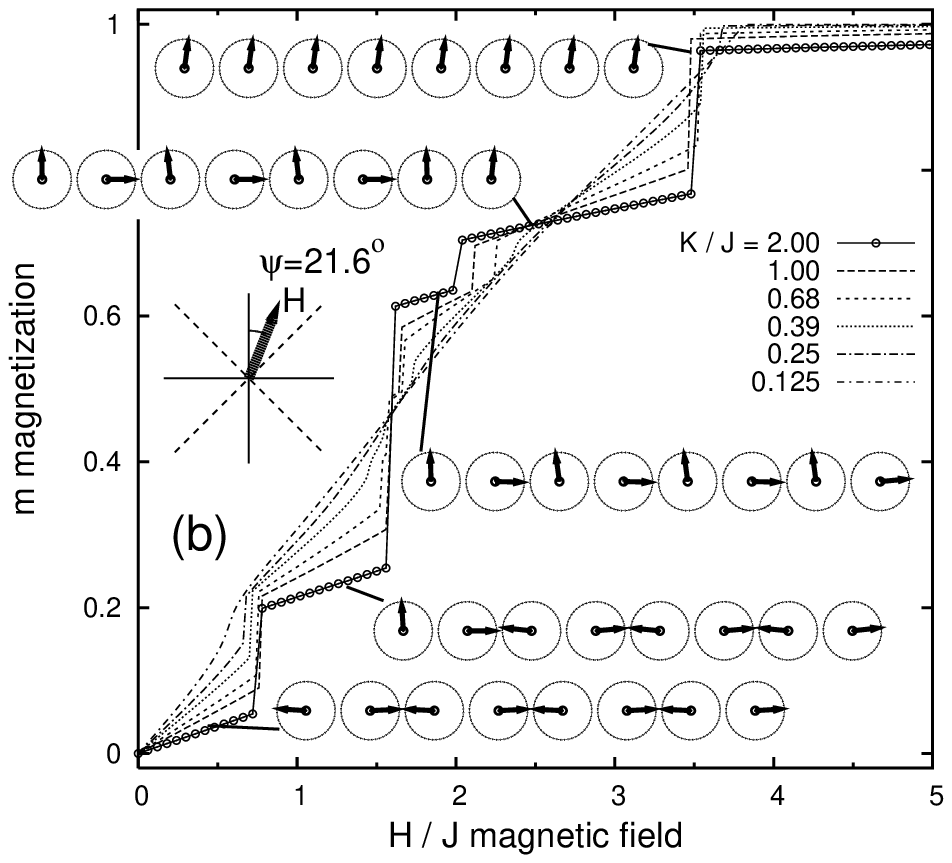}
\vspace{12pt}
\includegraphics[width=7.6cm]{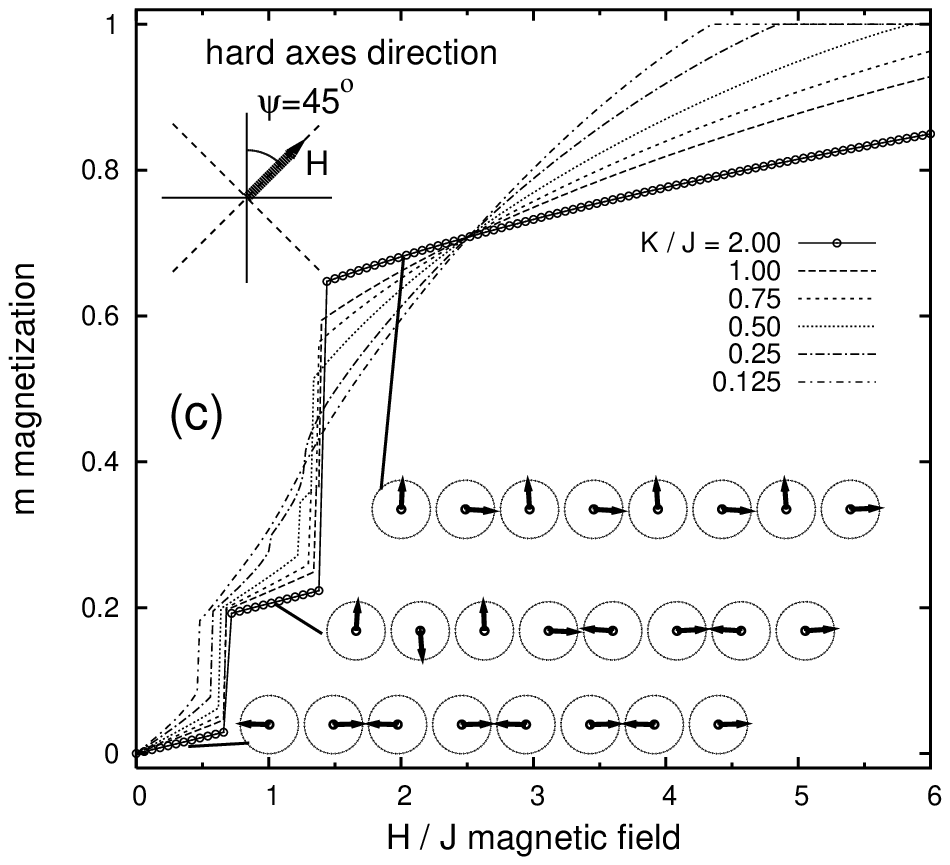}
\caption{
\label{MvsHN8}
Examples of magnetization in an antiferromagnetic
multilayer with $N=8$. 
Curves corresponding to the evolution of lowest energy 
states and for various values of anisotropy are shown 
with field oriented in directions of an easy axis (a),
in an oblique field (b) and in hard-axis direction (c).
$\psi$ is angle between magnetic field and easy axis direction.  
In each case, various magnetic 
configurations are shown for 
the discontinuous evolution of 
the magnetization curves
with highest anisotropy.
}
\end{figure}

For finite strength of the four-fold
anisotropy and under the influence of 
the exchange interactions, 
the basic structures 
are derived from the phases in Fig.~\ref{HxHyinfinite2}.
Under the influence of the field, they are elastically 
distorted into spatially \textit{inhomogeneous} configurations.
We have numerically investigated models for such 
cases with $N=4 \dots 20$ and various values 
of anisotropy $K$.
Fig.~\ref{MvsHN8} displays 
the general features
for the example $N=8$.
The numerically calculated
magnetization curves  corresponding
to the lowest energy states
for sufficiently high anisotropies
show the sequence of 
phases present in the infinite-anisotropy 
phase diagram Fig.~\ref{HxHyinfinite2}.
However, the degeneracy of these phases is lifted
because the distortions possible at finite anisotropy
yield different gains of energy for the different 
configurations of the phases L$^{(1)}$, F$^{(1)}$,
and L$^{(1)}$X$^{(1)}$.
Generally, depending on field-orientation, large domains 
with nearly spin-flop-like 
configurations $\mathbf{l} \perp \mathbf{H}$
are favoured because these configurations 
can be more easily distorted by the fields
yielding a corresponding gain of energy.
In particular, phases derived from 
the ferrimagnetic collinear F$^{(1)}$-phase 
are disfavoured compared to the L$^{(1)}$X$^{(1)}$-type phases.
The nearly collinear ferrimagnetic configurations 
seem to exist only as metastable states.
For fields closer to the hard-axis direction
the phases with either one L-pair or $N/2$
L-pairs occur.
Under the influence of an applied field 
further discontinuous transitions 
occur at intermediate anisotropies, 
as seen 
e.g. in Fig.~\ref{MvsHN8}~(a) for $K/J=0.375$.
These transitions are jumps
from one energy basin to another which are
formed by distorting degenerate configurations 
mainly of the type L$^{(1)}$ and L$^{(1)}$X$^{(1)}$.
%
%

At low anisotropies the phase
diagram attains the behaviour discussed
for zero anisotropies in section III.
Here the magnetic states are only 
influenced by the cuts of exchange bonds
at the surfaces and spatially
inhomogeneous spin-flop states slightly distorted
by four-fold anisotropy are realized (see Figs.~2 to 4).
($H_x,H_y$)-diagrams for this isotropic case consist
of the region of the inhomogeneous SF phase (Fig.~2, 3) 
separated from the saturated flip state 
by the critical line $H_e^{(N)}$ (Fig.~6).
These global features of the phase-diagrams
are similar to the case 
of antiferromagnetic multilayers 
with uniaxial anisotropy.\cite{PRB03}

Up to now we have discussed 
the effects of the competition between
the bilinear exchange and four-fold anisotropy. 
It is clear that finite \textit{biquadratic coupling}
may substantially change 
the critical fields and stability
regions of the various phases
as it affects the elastic stiffness of the system.
We note further, that the simple structure
of the phase-diagram 
ruled by the phases present in the infinite anisotropy
limit of Fig.~\ref{HxHyinfinite2} 
is valid only for the case of equal exchange constants $J_i \equiv J$
in the multilayer stack.
For arbitrary sets of values for $J_i$
in energy (\ref{energy1}) the phase diagrams 
may become considerably more complex and 
may contain further phases with different 
combinations of flipped 90 and 180 degree spin-pairs.
Even in such cases, 
the outline of the magnetic phase-diagrams,
described here for finite equal-constant superlattices,
should generally hold: 
(i) The high anisotropy limit is comparably
simple with few phases determined 
by the competing lowest-energy basins of the anisotropy.
Non-equal constants may lift some degeneracies 
that are present in superlattices.
(ii) In intermediate anisotropy range many elastically 
distorted phases appear, which are derived 
from stable and metastable high-anisotropy phases.
(iii) For vanishing anisotropy the phase-diagrams become
simple again as only inhomogeneous spin-flop-like phases
and the saturated ferromagnetic phase remain in external fields.

\subsection{Magnetization  processes in real systems}
So far, we have analyzed single domain magnetic 
configurations. 
Magnetization processes in real systems, however, 
are usually accompanied by complex reconstruction of 
the multidomain patterns as those recently observed 
in Fe/Cr multilayers.\cite{Nagy02, Lauter03}

There are two main physical mechanism for
multidomain states in the systems
under discussion.
Four-fold degeneracy of the ground-state
leads to creation of \textit{antiferromagnetic
multidomain structures} with 90 degree domain walls. 
Unlike the case of magnetic materials with nonzero
total magnetization, where multidomain
states are caused by demagnetization effects,
in antiferromagnets such domains are
\textit{metastable} and arise during the formation of the 
ordering state, i.e. they have a 
kinetic origin.\cite{Hubert74}
Hence, multidomain patterns observed
in antiferromagnetic coupled two-layers and
multilayers have irregular morphologies 
and depend on thermal and magnetic-field histories.\cite{Hubert98}

However, another type of multidomain structures 
arises in the vicinity of field-induced 
discontinuous transitions.\cite{UFN88}
Such thermodynamically stable 
\textit{ transitional domain structures} 
are formed by domains 
from states corresponding
to the coexisting phases at first-order transitions.
These domains are analogous to the domains of a
demagnetized ferromagnet. 
In principle, the equilibrium parameters
of such multidomain structures and their boundaries
can be calculated by standard methods.\cite{UFN88, Hubert98}
In \cite{FTT90} such calculations have been carried out
for bulk easy-plane tetragonal antiferromagnets.
\textit{Magnetoelastic interactions} lead to modification
of the inhomogeneous states and decrease the regions
of the multidomain states up 
to their complete suppression.\cite{Hubert98, UFN88, JMMM03}
This and other coercivity mechanism partly block
the development of the equilibrium states. 
As a result, in real systems the evolution of 
multidomain states \cite{Nagy02, Lauter03} 
is accompanied by rather strong hystereses.\cite{Zabel94}
For experiments, there are two important consequences
related to the starting states and history dependence
of magnetization processes. 
The ``texture'' of a real antiferromagnetic state 
in zero-field depends on the detailed kinetics 
imposed by e.g. cooling rates or deposition conditions.
On the other hand, field cycling by inner loops 
for an antiferromagnetic multilayer may yield various 
metastable configurations and multidomain structures
owing to the very wide coexistence region of the 
domains whenever sizable magnetic anisotropies are present 
in the multilayer stack.

\section{Conclusions}
Within a phenomenological approach
we develop the theory of reorientation
transition in antiferromagnetically 
coupled superlattices with in-plane magnetization.
Detailed investigations of the surface effects 
in the isotropic multilayers (Sec.~III)  and 
four-fold anisotropy effects in two-layer systems
(Sec.~IV) reveal the most important features of the system:
(i) Complex evolution of the inhomogeneous
states (Figs.~2-4) is imposed by the strong disbalance
of the exchange coupling, i.e. by the 
\textit{cut of the exchange bonds}.
(ii) Remarkable field-induced reorientational processes 
occur due to enhanced four-fold anisotropy 
(Figs.~8 to \ref{MvsHN8}).

The model used here, corresponding solutions and phase-diagrams 
include as special cases earlier theoretical studies 
on surface \cite{Nortemann92} and four-fold 
anisotropy \cite{Dieny90, FTT90} effects.
The results are in essential accordance with  
existing experimental observations on inhomogeneous
distribution in the spin-flop phase near the saturation
field  \cite{Lauter02},
and some effects of four-fold anisotropy in
Co/Cu(001) wedged two-layers \cite{Zabel94} and 
Fe/Cr superlattices.\cite{ Ustinov01}
Our approach and results enable a \textit{qualitative} 
analysis of the magnetization processes 
in the multilayered systems (Sec. V). 
In spite of the rather complex 
phase-diagrams of these systems, 
the analysis can be extended towards 
a quantitative description of real systems 
belonging to the class of artificial layered antiferromagnets 
described by Eq.~(\ref{energy1}).

So far most experimental work is 
carried out only for special conditions,
often data are collected only with fields along easy-axes.
These results only cover 
small regions of the ($H_x$,$H_y$)-phase diagrams 
(Figs.~\ref{HxHyweak} to \ref{HxHyinfinite2}) 
and do not capture the rich varieties of
magnetizations processes in such systems.
It is remarkable that many interesting effects ,
such as 90-degree folded phases, transitions into
asymmetric canted phases etc.,
are present already in antiferromagnetically
coupled two-layers.\cite{Zabel94}
As we have seen, phase diagrams for multilayer 
systems with $N>2$ become very complex.
Hence, systems with few layers 
are probably a better starting point 
for detailed investigations on magnetization processes.
Such experiments could be used to assess 
magnetic parameters and quality of such systems.
%
%
%
Generally, magnetization processes and checks 
for their dependence on magnetic and thermal pre-history
should be made by applying fields 
in oblique directions and/or under rotating fields.
Only then, the behaviour of the ($H_x$,$H_y$)-phase 
diagrams can be usefully compared with 
detailed theoretical investigations. 

Further theoretical work may 
address models with non-equal constants.
Also effectively ferrimagnetic systems
with odd numbers $N$ of layers 
and with different layer thicknesses
may be interesting.
Some experimental data for magnetization processes 
in such systems exist, e.g. for  Co/Cr two-layers
with different ferromagnetic layers thicknesses,\cite{Hilt99}
and for odd-numbered multilayers.\cite{Zabel94}
Interesting reorientational effects should arise also
in magnetic fields, which are applied perpendicular 
or under arbitrary angle to the layer plane.
Antiferromagnetically coupled superlattices may
also undergo transitions into perpendicularly 
magnetized states for certain thicknesses
of the individual ferromagnetic layers,
as observed for Co/Cr(001).\cite{Zabel96}
In such cases the stray field must be taken into account
already for laterally homogeneous states.

Concluding we state that the magnetic effects
and phenomena discussed in this paper 
can be used for detailed investigations 
on such aspects of nanomagnetism 
as interlayer-exchange interactions, 
reorientational, and multidomain processes.

\begin{acknowledgments}
A.\ N.\ B.\ thanks H.\ Eschrig for support and
hospitality at IFW Dresden. 
We thank S{\"a}chsisches Staatsministerium f{\"u}r Wissenschaft 
und Kunst for financial support.
\end{acknowledgments}


\end{document}